\documentclass[aps,prb,superscriptaddress,showpacs,floatfix]{revtex4-1}
\usepackage{amsfonts}
\usepackage{graphicx,amsmath,amssymb,xspace,epsfig,float,multirow,subfigure,tabularx}

\pdfoutput=1
\begin{document}
\title{Strong magnetic field induces superconductivity in Weyl semi - metal.}
\author{Baruch Rosenstein}
\email{baruchro@hotmail.com}
\affiliation{Electrophysics Department, National Chiao Tung University, Hsinchu 30050,
\textit{Taiwan, R. O. C}}
\author{B.Ya. Shapiro}
\email{shapib@mail.biu.ac.il}
\affiliation{Physics Department, Bar-Ilan University, 52900 Ramat-Gan, Israel}
\author{Dingping Li}
\email{lidp@pku.edu.cn}
\affiliation{School of Physics, Peking University, Beijing 100871, \textit{China}}
\affiliation{Collaborative Innovation Center of Quantum Matter, Beijing, China}
\author{I. Shapiro}
\affiliation{Physics Department, Bar-Ilan University, 52900 Ramat-Gan, Israel}
\date{\today}

\begin{abstract}
Microscopic theory of the normal-to-superconductor coexistence line of a
multi-band Weyl superconductor subjected to magnetic field is constructed.
It is shown that Weyl semi-metal that is nonsuperconducting or having a
small critical temperature $T_{c}$ at zero field, might become
superconductor at higher temperature, when the magnetic field is tuned to a
series of quantized values $H_{n}$. The pairing occurs on Landau levels. It
is argued that the phenomenon is much easier detectable in Weyl semi -
metals than in parabolic band metals since the quantum limit already has
been approaches in several Weyl materials. The effect of Zeeman coupling
leading to splitting of the reentrant superconducting regions on the
magnetic phase diagram is considered. An experimental signature of the
superconductivity on Landau levels is reduction of magnetoresistivity. This
has already been observed in $Cd_{3}As_{2}$ and several other compounds. The
novel kind of quantum oscillations of magnetoresistance detected in $%
ZrTe_{5} $ is discussed along these lines.
\end{abstract}


\maketitle

\section{Introduction}

Conventional superconductivity arises from pairing of electrons in the
vicinity of the Fermi surface, since the phonon mediated attraction is
effective only when the electron's energy is within a shell of the Debye
energy width, $\hbar \Omega $ of order several hundreds of kelvin, see
Fig.1. Within the BCS theory (in the adiabatic limit) the order parameter, $%
\Delta \sim T_{c}$, depends exponentially on the density of states (DOS) at
Fermi level $D\left( \mu \right) $, so that in order to enhance the tendency
for superconductivity, one should use any means to boost the density of
states within this narrow shell. In quantum systems there is an obvious way
to boost locally the DOS - quantization. Thus a natural mean to concentrate
the spectral weight is a strong magnetic field that causes Landau
quantization. The best known example of this phenomenon is 2D the electron
gas in magnetic field, where DOS can be tuned to "infinity" at certain
values of magnetic fields and the quantum Hall effect became visible.

In principle, one can imagine that strong magnetic field can enhance
superconductivity as well, if the quantum limit (when the Fermi surface
crosses the lowest Landau levels) is reached. At first glance there are two
immediate problems with this scenario. First the magnetic field generally
breaks the Cooper pairs due to the orbital instability that leads\cite%
{Hochenberg} to suppression of superconductivity at $H_{c2}$. Second, the
direct (Zeeman) coupling of the magnetic field to the electron's spin also
leads (for the singlet pairing) to the Chandrasekhar - Klogston\cite%
{Chandrasekhar} pair breaking at $H_{p}$. However it was predicted in
eighties of the last century (see \cite{Indians,Maniv,RMP} and references
therein) that paradoxically superconductivity can reappear on the Landau
levels (LL) at fields far above $H_{c2}$. While the superconductivity
enhancement can occur at any LL, it is stable against perturbations only
near the "quantum limit", in which the lowest LL level crosses the Fermi
energy $\mu $. \ The condition for that, $\mu \sim \hbar \omega _{c}^{p}$,
however restricts the choice of material to those with extremely small
electron density. Even for $100T$ the Fermi level should be just $10meV$.

In conventional metallic superconductors, even at $H_{c2}=\Phi _{0}/2\pi \xi
^{2}$ (where $\xi $ is the coherence length at zero temperature and $\Phi
_{0}$ is the flux quantum), the effect of the Landau quantization of the
electron motion is negligible. For a metal with effective mass $m^{\ast }$,
the separation between (equidistant) Landau levels is $\hbar \omega
_{c}^{p}=\hbar eH/m^{\ast }c$. For typical values of the field $H_{c2}=3T$
and effective mass $m^{\ast }\sim m_{e}$, the level spacing is $4K$, much
smaller than $2\hbar \Omega $. Therefore, to take advantage of the Landau
quantization effect on superconductivity, one should consider a super strong
magnetic field of thousands Tesla. The estimate however is based on the
assumption of the parabolic dispersion relation of the normal electrons (or\
holes).

Recently a new class of 2D and 3D multi-band materials with qualitatively
different band structure near the Fermi level was discovered\cite%
{Geim,superWSM2D,superWSM3D,Cao, ZrTe2,Hf} -\ Weyl (Dirac) semi-metals
(WSM). Unlike in conventional semi-metals with several quasiparticle and
hole bands, in WSM Dirac points occur due to the band inversion near the
Fermi level. WSM are characterized by linear dispersion relation, $%
\varepsilon =vp$, and in many of them the chemical potential is tunable and
small. Even a more important fact for pairing is that their inter - band
tunneling is dominant. In some of this novel materials conventional phonon
mediated superconductivity with $T_{c}$ up to $20K$ (under pressure) with $%
H_{c2}$ of several $T$ was achieved \cite{superWSM2D,superWSM3D}. Although
mechanism of superconductivity is these materials does not differ much from
the low $T_{c}$ metals\cite{DasSarma,JCM15}, the position of the Landau
levels (LL) does. The notion of the effective mass does not apply for this
essentially non-parabolic dispersion relation and LL are generally no longer
equidistant\cite{Geim}, see Fig.1. This raises a possibility that the Landau
quantum limit is easier achievable in this case\cite{Cao}. The first LL
appears at $\hbar \omega _{c}=v\sqrt{2\hbar eH/c}$ should be equal to $\mu $
counted from the Dirac point. For a typical values of $v=10^{8}cm/s$ and $%
H=100T$, now one obtains $\mu =0.4eV$, that favorably compares with the
previous estimate of $10meV$ in a "conventional" parabolic band. The
condition for the superconductivity enhancement in WSM is thus qualitatively
different and quantum limit condition becomes $\omega _{c}\hbar \sim 2\hbar
\Omega $. A more quantitative estimates and comparison between the
conventional materials and the WSM is made below.
\begin{figure}[tbp]
\begin{center}
\includegraphics[width=8cm]{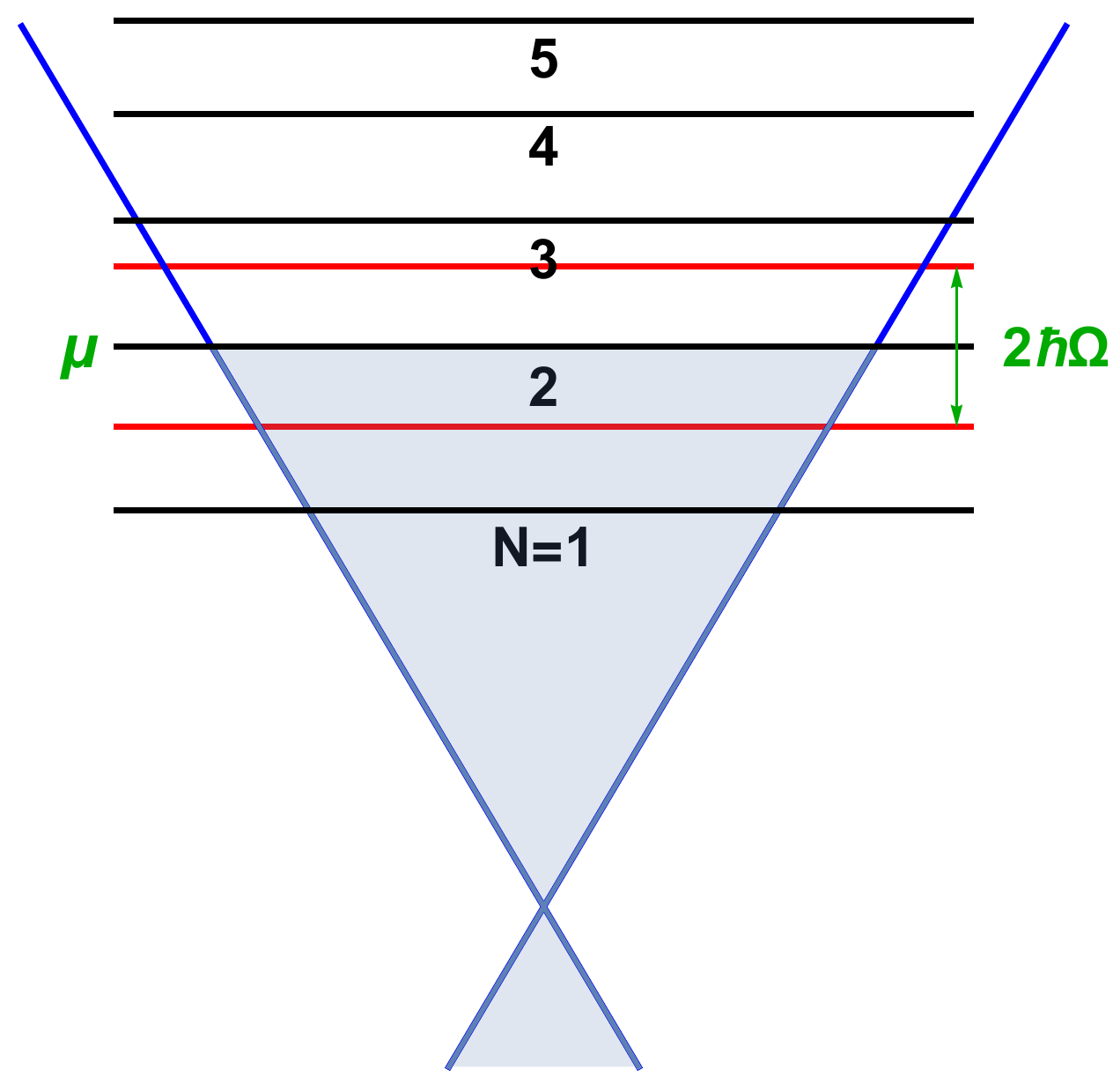}
\end{center}
\par
\vspace{-0.5cm}
\caption{Set of Landau levels in Weyl semimetals. Pairing due to phonons
occurs in the energy shell of Debye energy width, $\hbar \Omega $, around
the Fermi level $\protect \mu $.}
\end{figure}
Therefore it is important to extend the BCS type theory to the case of multi
- band semi - metals like the WSM. The extension of conventional Gor'kov-
Eliashberg approach in strong magnetic field\cite{Indians,RMP,Maniv} to a
multi - band semi -metals by no means trivial. For two parabolic (one
quasi-particle and one hole) bands it was done in ref.\onlinecite{Koshelev}.
Since in WSM ratio $\mu /\hbar \Omega $ is relatively small, an important
additional issue is the role of the retardation effects of the phonon
mediated pairing in order remain within the bounds of the adiabatic
approximation.

In this paper the effect of the phonon - mediated pairing in strong magnetic
fields (including the quantum limit) in Weyl semi-metals is developed in
wide range of temperatures and magnetic fields. The simplest model
necessarily contains four (sub) bands (two Weyl subbands and two
magnetically split spin subbands due to Zeeman coupling). The magnetic phase
diagram consist of a series of superconducting domes in addition to the
conventional $H_{c2}\left( T\right) $ line. Recent experiments \cite{Cao} on
$Cd_{3}As_{2}$ in fields up to $52T$ are reinterpreted as possible candidate
of re-entrant superconductivity at $N=2,3$ Landau levels at $25T$ and $46T$.
It is interesting to note that the upper bound on superconductivity at zero
field in this material is $3K$. Retardation effects of the phonon mediated
pairing is discussed and taken into account phenomenologically.

The paper is organized as follows. The effect of re-entrant
superconductivity at very high magnetic fields is more pronounced in two
dimensions, so a sufficiently general 2D WSM model is defined in Section II.
The superconductor-normal phase transition line in 2D WSM in high magnetic
fields is derived in Section III. The phase diagram of superconductivity on
Landau levels is extended to Zeeman coupling and to the anisotropic 3D WSM
in Section IV. Comparison with recent experiments, discussion and
conclusions is the subject of Section V.

\section{Phonon mediated superconductivity in WSM in strong magnetic field.}

\subsection{Pairing in 2D WSM under magnetic field}

A Weyl material typically possesses several sublattices. We exemplify the
effect of the WSM band structure on superconductivity using the simplest
model with just two sublattices denoted by $\alpha =1,2$. The effective
electron-electron attraction due to the electron - phonon coupling overcomes
the Coulomb repulsion and induces pairing. Typically in WSM there are
numerous bands. We assume that different valleys are paired independently
and drop all the valley indices (including chirality, multiplying the
density of states by $2N_{f}$). To simplify notations, we therefore consider
just one spinor (left, for definiteness), the following Weyl Hamiltonian\cite%
{Wang},\cite{JCM15}.%
\begin{equation}
K=\int_{\mathbf{r}}\psi _{\alpha }^{s\dagger }\left( \mathbf{r}\right) \left
\{ -i\hbar v\left( D_{x}\sigma _{\alpha \beta }^{x}+D_{y}\sigma _{\alpha
\beta }^{y}\right) -\mu \delta _{\alpha \beta }\right \} \psi _{\beta
}^{s}\left( \mathbf{r}\right) .  \label{K}
\end{equation}%
Here $v$ is Fermi velocity assumed isotropic in the plane $x-y$
perpendicular to the applied magnetic field (assumed isotropic, generalized
later to anisotropic 3D WSM). Chemical potential is denoted by $\mu $ -
chemical potential. Pauli matrices $\sigma $ operate in the sublattice space
(the indices $\alpha ,\beta $ will be termed the pseudo-spin projections)
and $s$ is spin projection. Magnetic field appears in the covariant
derivatives via the vector potential, $D_{i}=\nabla ^{i}-i\frac{e}{\hbar c}%
A_{i}$. Here $\mathbf{A}$ is the vector potential.

Further we assume the local density - density interaction Hamiltonian \cite%
{Abrikosov},
\begin{equation}
V=\frac{g^{2}}{2}\int_{\mathbf{r}}\psi _{\alpha }^{+\uparrow }\left( \mathbf{%
r}\right) \psi _{\alpha }^{\downarrow }\left( \mathbf{r}\right) \psi _{\beta
}^{\downarrow +}\left( \mathbf{r}\right) \psi _{\beta }^{\uparrow }\left(
\mathbf{r}\right) \text{,}  \label{int}
\end{equation}%
ignoring the Coulomb repulsion (that as usual is accounted for by a
pseudopotential, so that $g$ is the electron - phonon coupling). It is
important that the interaction has a cutoff Debye frequency $\Omega $, so
that it is active in an energy shell of width $2\hbar \Omega $ around the
Fermi level \cite{Abrikosov}. We will discuss a more realistic dependence on
frequency in Section III.

\subsection{ Matsubara Green's functions and Gor'kov equations.}

Finite temperature properties of the superconducting condensate are
described by the normal and the anomalous Matsubara Green's functions\cite%
{Abrikosov} (GF),%
\begin{eqnarray}
G_{\alpha \beta }^{ts}\left( \mathbf{r}\tau ,\mathbf{r}^{\prime }\tau
^{\prime }\right) &=&-\left \langle T\psi _{\alpha }^{t}\left( \mathbf{r}%
\tau \right) \psi _{\beta }^{\dagger s}\left( \mathbf{r}^{\prime }\tau
^{\prime }\right) \right \rangle ;F_{\alpha \beta }^{ts}\left( \mathbf{r}%
\tau ,\mathbf{r}^{\prime }\tau ^{\prime }\right) =\left \langle T\psi
_{\alpha }^{t}\left( \mathbf{r}\tau \right) \psi _{\beta }^{s}\left( \mathbf{%
r}^{\prime }\tau ^{\prime }\right) \right \rangle ;  \label{GF} \\
F_{\alpha \beta }^{+ts}\left( \mathbf{r}\tau ,\mathbf{r}^{\prime }\tau
^{\prime }\right) &=&\left \langle T\psi _{\alpha }^{\dagger t}\left(
\mathbf{r}\tau \right) \psi _{\beta }^{\dagger s}\left( \mathbf{r}^{\prime
}\tau ^{\prime }\right) \right \rangle ,  \notag
\end{eqnarray}%
with the spin Ansatz
\begin{eqnarray}
G_{\alpha \beta }^{ts}\left( \mathbf{r}\tau ,\mathbf{r}^{\prime }\tau
^{\prime }\right) &=&\delta ^{ts}G_{\alpha \beta }\left( \mathbf{r,r}%
^{\prime },\tau -\tau ^{\prime }\right) ;F_{\alpha \beta }^{ts}\left(
\mathbf{r}\tau ,\mathbf{r}^{\prime }\tau ^{\prime }\right) =-\varepsilon
^{ts}F_{\alpha \beta }\left( \mathbf{r,r}^{\prime },\tau -\tau ^{\prime
}\right) ;  \label{spin Ansatz} \\
F_{\alpha \beta }^{+ts}\left( \mathbf{r}\tau ,\mathbf{r}^{\prime }\tau
^{\prime }\right) &=&\varepsilon ^{ts}F_{\alpha \beta }^{+}\left( \mathbf{r,r%
}^{\prime },\tau -\tau ^{\prime }\right) \text{.}  \notag
\end{eqnarray}
Here the Plank constant is set to $\hbar =1$. Using the Fourier transform,
\begin{equation}
G_{\gamma \kappa }\left( \mathbf{r},\tau \right) =T\sum \nolimits_{s}\exp %
\left[ -i\omega _{s}\tau \right] G_{\gamma \kappa }\left( \omega ,\mathbf{r}%
\right) \text{,}  \label{Fourier2D}
\end{equation}%
with fermionic Matsubara frequencies, $\omega _{s}=2\pi T\left( s+1/2\right)
$,\ one obtains from equations of operator motion the set of Gor'kov
equations, see ref. \onlinecite{Rosenstein} generalized to include magnetic
field:
\begin{eqnarray}
i\omega G_{\gamma \kappa }\left( \mathbf{r,r}^{\prime },\omega \right) +i\
vD_{\mathbf{r}}^{i}\sigma _{\gamma \beta }^{i}G_{\beta \kappa }\left(
\mathbf{r,r}^{\prime },\omega \right) +\mu G_{\gamma \kappa }\left( \mathbf{%
r,r}^{\prime },\omega \right) +\Delta _{\alpha \gamma }\left( \mathbf{r,}%
0\right) F_{\alpha \kappa }^{+}\left( \mathbf{r,r}^{\prime },\omega \right)
&=&\delta ^{\gamma \kappa }\delta \left( \mathbf{r-r}^{\prime }\right) ;
\label{GEqua} \\
-i\omega F_{\gamma \kappa }^{+}\left( \mathbf{r},\mathbf{r}^{\prime },\omega
\right) -ivD_{\mathbf{r}}^{i}\sigma _{\alpha \gamma }^{i}F_{\alpha \kappa
}^{+}\left( \mathbf{r},\mathbf{r}^{\prime },\omega \right) +\mu F_{\gamma
\kappa }^{+}\left( \mathbf{r},\mathbf{r}^{\prime },\omega \right) -\Delta
_{\alpha \gamma }^{\ast }\left( \mathbf{r},0\right) G_{\alpha \kappa }\left(
\mathbf{r},\mathbf{r}^{\prime },\omega \right) &=&0\text{.}  \notag
\end{eqnarray}%
It will be shown that the singlet pairing pseudo-spin Ansatz, $\Delta
_{\alpha \gamma }\equiv \sigma _{\alpha \gamma }^{x}\Delta $, obeys the
Pauli principle. The gap function consequently reads: $\Delta =\frac{1}{2}Tr%
\left[ \sigma ^{x}\widehat{\Delta }\right] $. Notice, that in contrast to
conventional metals with parabolic dispersion law, in the case of the Weyl
semi - metals the second Gor'kov equation, Eq.(\ref{GEqua}), contains
transposed Pauli matrices for isospins.

\section{The transition line}

In this Section the superconductor-normal phase transition line in high
magnetic fields is determined. The line breaks into a set of disconnected
segments since in certain cases the superconductivity reappears when a
Landau level crosses Fermi surface.

\subsection{Linearization of the Gor'kov equations near the transition line}

Near the normal-to-superconducting transition line the gap $\Delta $ is
small and the set of the Gor'kov equations \ref{GEqua} can be linearized. In
this case the gap equation describing the critical curve $H_{c2}\left(
T\right) $ has \ the form, see ref. \onlinecite{Rosenstein} for details,
\begin{eqnarray}
\Delta \left( \mathbf{r}\right) &=&\frac{g^{2}}{2}T\sum \nolimits_{\omega
}\int_{\mathbf{r}^{\prime }}\Delta ^{\ast }\left( \mathbf{r}^{\prime
}\right) \sigma _{\kappa \beta }^{x}G_{\beta \gamma }^{2}\left( \mathbf{%
r^{\prime },r}\right) \sigma _{\gamma \alpha }^{x}G_{\alpha \kappa
}^{1}\left( \mathbf{r,r}^{\prime }\right)  \label{EqH} \\
&=&\frac{g^{2}}{2}\sum \nolimits_{\omega }\int_{\mathbf{r}^{\prime }}\Delta
^{\ast }\left( \mathbf{r}^{\prime }\right) \left(
\begin{array}{c}
G_{2\mathbf{2}}^{2}\left( \mathbf{r^{\prime },r}\right) G_{1\mathbf{1}%
}^{1}\left( \mathbf{r,r}^{\prime }\right) +G_{1\mathbf{1}}^{2}\left( \mathbf{%
r^{\prime },r}\right) G_{2\mathbf{2}}^{1}\left( \mathbf{r,r}^{\prime }\right)
\\
+G_{12}^{2}\left( \mathbf{r^{\prime },r}\right) G_{12}^{1}\left( \mathbf{r,r}%
^{\prime }\right) +G_{2\mathbf{1}}^{2}\left( \mathbf{r^{\prime },r}\right)
G_{2\mathbf{1}}^{1}\left( \mathbf{r,r}^{\prime }\right)%
\end{array}%
\right) \text{.}  \notag
\end{eqnarray}%
Here the normal GF is obtained from,$\ $%
\begin{equation}
\left[ \ iv\mathbf{D}_{\mathbf{r}}\cdot \mathbf{\sigma }_{\gamma \beta
}+\left( i\omega +\mu \right) \delta _{\gamma \beta }\right] G_{\beta \kappa
}^{1}\left( \mathbf{r,r}^{\prime }\right) =\delta ^{\gamma \kappa }\delta
\left( \mathbf{r-r}^{\prime }\right) \text{,}  \label{normalGF}
\end{equation}%
while a quantity$\  \overline{G}_{\beta \gamma }$ (an auxiliary function
associated with $G$ via a product of an axis reflection and time reversal)
obeys a \textit{different} equations:
\begin{equation}
\left[ -iv\mathbf{D}_{\mathbf{r}}\cdot \mathbf{\sigma }_{\gamma \beta
}^{t}+\left( -i\omega +\mu \right) \delta _{\gamma \beta }\right] G_{\beta
\kappa }^{2}\left( \mathbf{r}^{\prime }\mathbf{,r}\right) =\delta ^{\gamma
\kappa }\delta \left( \mathbf{r-r}^{\prime }\right) \text{.}  \label{n2a}
\end{equation}%
Here $\mathbf{\sigma }^{t}$ is the transposed Pauli matrix that replaces $%
\mathbf{\sigma }$ in the customary normal state equation Eq.(\ref{normalGF}).

In the uniform magnetic field the GF can be written (in the symmetric gauge,
$\mathbf{A=}\frac{1}{2}\mathbf{H\times r}$) in the following form:%
\begin{eqnarray}
G_{\beta \kappa }^{1}\left( \mathbf{r},\mathbf{r}^{\prime }\right) &=&\exp %
\left[ -i\frac{xy^{\prime }-yx^{\prime }}{2l^{2}}\right] g_{\beta \kappa
}^{1}\left( \mathbf{r-r}^{\prime }\right) ;  \label{n3a} \\
G_{\beta \kappa }^{2}\left( \mathbf{r}^{\prime },\mathbf{r}\right) &=&\exp %
\left[ -i\frac{xy^{\prime }-yx^{\prime }}{2l^{2}}\right] g_{\beta \kappa
}^{2}\left( \mathbf{r}^{\prime }\mathbf{-r}\right) \text{.}  \notag
\end{eqnarray}%
Here $l^{2}=c/eH$ is the magnetic length. This phase Ansatz indeed works.
Substituting it into Eq.(\ref{normalGF}) and Eq.(\ref{n2a}) respectively,
the variables separate:

\begin{equation}
\left \{ \left( i\omega +\mu \right) \delta _{\gamma \beta }-v\mathbf{\Pi
\cdot \sigma }_{\gamma \beta }\  \right \} g_{\beta \kappa }^{1}\left(
\mathbf{r-r}^{\prime }\right) =\delta ^{\gamma \kappa }\delta \left( \mathbf{%
r-r}^{\prime }\right) ;  \label{normalg1}
\end{equation}%
\begin{equation}
\left \{ \left( -i\omega +\mu \right) \delta _{\gamma \beta }+v\mathbf{\Pi }%
\cdot \mathbf{\sigma }_{\gamma \beta }^{t}\right \} g_{\beta \kappa
}^{2}\left( \mathbf{r}^{\prime }\mathbf{-r}\right) =\delta ^{\gamma \kappa
}\delta \left( \mathbf{r-r}^{\prime }\right) .  \label{n4ac}
\end{equation}%
Here the ladder operators here are defined as
\begin{equation}
\Pi _{x}=\mathbf{-}i\frac{\partial }{\partial \rho _{x}}+\frac{\ 1}{2l^{2}}%
\rho _{y},\Pi _{y}=-i\frac{\partial }{\partial \rho _{y}}-\frac{\ 1}{2l^{2}}%
\rho _{x},  \label{Ladder}
\end{equation}%
with relative distance denoted by $\mathbf{\rho =r-r}^{\prime }$.

These equations are solved by expansion in the basis of eigenfunctions of
harmonic oscillator in Appendix A. The resulting normal GF in terms of
generalized Laguerre polynomials are:
\begin{eqnarray}
g_{11}^{1}\left( \mathbf{\rho }\right) &=&\frac{\left( i\omega +\mu \right)
}{2\pi l^{2}}\exp \left[ -\frac{\rho ^{2}}{4l^{2}}\right] \sum_{n=0}\frac{\
L_{n}\left[ \rho ^{2}/2l^{2}\right] }{\left( i\omega +\mu \right)
^{2}-\omega _{c}^{2}\left( n+1\right) };  \label{GF1} \\
g_{21}^{1}\left( \mathbf{\rho }\right) &=&-\  \frac{iv\rho e^{i\theta }}{2\pi
l^{4}}\exp \left[ -\frac{\rho ^{2}}{4l^{2}}\right] \sum_{n=1}\frac{%
L_{n-1}^{1}\left[ \rho ^{2}/2l^{2}\right] }{\left( i\omega +\mu \right)
^{2}-\omega _{c}^{2}\left( n+1\right) };  \notag \\
g_{22}^{1}\left( \mathbf{\rho }\right) &=&\frac{\left( i\omega +\mu \right)
}{2\pi l^{2}}\exp \left[ -\frac{\rho ^{2}}{4l^{2}}\right] \sum_{n=0}\frac{\
L_{n}\left[ \rho ^{2}/2l^{2}\right] }{\left( i\omega +\mu \right)
^{2}-\omega _{c}^{2}n};  \notag \\
g_{12}^{1}\left( \mathbf{\rho }\right) &=&-\  \frac{iv\rho e^{-i\theta }}{%
2\pi l^{4}}\exp \left[ -\frac{\rho ^{2}}{4l^{2}}\right] \sum_{n=1}\frac{%
L_{n}^{1}\left[ \rho ^{2}/2l^{2}\right] }{\left( i\omega +\mu \right)
^{2}-\omega _{c}^{2}n\ }\text{.}  \notag
\end{eqnarray}%
Here the cyclotron frequency in WSM is denoted by $\omega _{c}^{2}=$ $%
2v^{2}/l^{2}$ and $\theta $ is the polar angle of $\mathbf{\rho }$.
Similarly the associate GF are:
\begin{eqnarray}
g_{11}^{2}\left( -\mathbf{\rho }\right) &=&\frac{-i\omega +\mu }{2\pi l^{2}}%
\exp \left[ -\frac{\rho ^{2}}{4l^{2}}\right] \sum_{n=0}\frac{L_{n}\left[
\rho ^{2}/2l^{2}\right] }{\left( -i\omega +\mu \right) ^{2}-\omega _{c}^{2}n}%
;  \label{GF2} \\
g_{12}^{2}\left( -\mathbf{\rho }\right) &=&\frac{iv\rho e^{i\theta }}{2\pi
l^{4}}\exp \left[ -\frac{\rho ^{2}}{4l^{2}}\right] \sum_{n=1}\frac{%
L_{n-1}^{1}\left[ \rho ^{2}/2l^{2}\right] }{\left( -i\omega +\mu \right)
^{2}-\omega _{c}^{2}\left( n+1\right) };  \notag \\
g_{21}^{2}\left( -\mathbf{\rho }\right) &=&\frac{iv\rho e^{-i\theta }}{2\pi
l^{4}}\exp \left[ -\frac{\rho ^{2}}{4l^{2}}\right] \sum_{n=1}\frac{L_{n}^{1}%
\left[ \rho ^{2}/2l^{2}\right] }{\left( -i\omega +\mu \right) ^{2}-\omega
_{c}^{2}n};  \notag \\
g_{22}^{2}\left( -\mathbf{\rho }\right) &=&\frac{-i\omega +\mu }{2\pi l^{2}}%
\exp \left[ -\frac{\rho ^{2}}{4l^{2}}\right] \sum_{n=0}\frac{L_{n}\left[
\rho ^{2}/2l^{2}\right] }{\left( -i\omega +\mu \right) ^{2}-\omega
_{c}^{2}\left( n+1\right) }\text{.}  \notag
\end{eqnarray}%
Now we are ready to return to the gap equation at criticality.

\subsection{Ansatz for the gap function and the angle integration}

Substituting the phase factors of GF from Eq.(\ref{n3a}) into the gap
equation, Eq.(\ref{EqH}), one obtains:
\begin{equation}
\Delta \left( \mathbf{r}\right) =\frac{g^{2}T}{2}\sum \nolimits_{\omega
}\int_{\mathbf{r}^{\prime }}\exp \left[ -i\frac{xy^{\prime }-yx^{\prime }}{%
l^{2}}\right] \Delta ^{\ast }\left( \mathbf{r}^{\prime }\right) \left(
\begin{array}{c}
g_{22}^{2}\left( \mathbf{-\rho }\right) g_{11}^{1}\left( \mathbf{\rho }%
\right) +g_{11}^{2}\left( \mathbf{-\rho }\right) g_{22}^{1}\left( \mathbf{%
\rho }\right) \\
+g_{12}^{2}\left( \mathbf{-\rho }\right) g_{12}^{1}\left( \mathbf{\rho }%
\right) +g_{21}^{2}\left( \mathbf{-\rho }\right) g_{21}^{1}\left( \mathbf{%
\rho }\right)%
\end{array}%
\right) \text{.}  \label{gap H}
\end{equation}%
Adopting the gaussian Ansatz for the gap function,
\begin{equation}
\Delta \left( \mathbf{r}\right) =\exp \left[ -r^{2}/2l^{2}\right] ,
\label{deltaAnsatz}
\end{equation}%
used extensively in calculations since the seminal work \cite{Hochenberg},
and substituting the above explicit expressions for the GF, one obtains,
\begin{equation}
1=\frac{g^{2}T}{8\pi ^{2}l^{4}}\sum \nolimits_{\omega }\int_{0}^{\infty
}\rho d\rho \int_{\theta =0}^{2\pi }\exp \left[ \frac{r\rho }{l^{2}}%
e^{i\theta }\right] \exp \left[ -2u\right] S\left( u,\omega \right) \text{,}
\label{kkk}
\end{equation}%
where the integral have been shifted to $\mathbf{\rho =r-r}^{\prime }$. The
scalar function $S$ depends on absolute value of $\mathbf{\rho }$ only, so
that the dimensionless variable $u=\rho ^{2}/2l^{2}$ is used instead. It is
a double sum over Landau levels:
\begin{equation}
S\left( u,\omega \right) =%
\begin{array}{c}
\left( \omega ^{2}+\mu ^{2}\right) \sum_{n,m=0}^{\infty }\left \{ \frac{L_{n}%
\left[ u\right] L_{m}\left[ u\right] }{\left( \left( -i\omega +\mu \right)
^{2}-\omega _{c}^{2}\left( n+1\right) \right) \left( \left( i\omega +\mu
\right) ^{2}-\omega _{c}^{2}\left( m+1\right) \right) }+\frac{L_{n}\left[ u%
\right] L_{m}\left[ u\right] }{\left( \left( -i\omega +\mu \right)
^{2}-\omega _{c}^{2}n\right) \left( \left( i\omega +\mu \right) ^{2}-\omega
_{c}^{2}m\right) }\right \} \\
+\omega _{c}^{2}\sum_{n,m=1}^{\infty }\left \{ \frac{uL_{n-1}^{1}\left[ u%
\right] L_{m}^{1}\left[ u\right] }{\left( \left( -i\omega +\mu \right)
^{2}-\omega _{c}^{2}\left( n+1\right) \right) \left( \left( i\omega +\mu
\right) ^{2}-\omega _{c}^{2}m\  \right) }+\frac{uL_{n}^{1}\left[ u\right]
L_{m-1}^{1}\left[ u\right] }{\left( \left( -i\omega +\mu \right) ^{2}-\omega
_{c}^{2}n\right) \left( \left( i\omega +\mu \right) ^{2}-\omega
_{c}^{2}\left( m+1\right) \right) }\right \} \text{.}%
\end{array}%
,  \label{s}
\end{equation}
The integral over $\theta $ is just\cite{Gradshtein} $2\pi $, so that the
gap equation at criticality takes a form$\ $
\begin{equation}
1=\frac{g^{2}T}{4\pi l^{2}}\sum \nolimits_{\omega }\int_{u=0}^{\infty }\exp %
\left[ -2u\right] S\left( u,\omega \right) \text{.}  \label{gapeqangle}
\end{equation}%
In what follows the integral over $u$ and the sum over the Matsubara
frequencies is explicitly performed and the equation used to investigate the
effect of Landau quantization of superconductivity in a WSM. Using the
integrals over product of generalized Laguerre polynomials\cite{Gradshtein},
\begin{eqnarray}
\int \limits_{0}^{\infty }du\exp \left( -2u\right) L_{n}\left( u\right)
L_{m}\left( u\right) &=&\frac{\left( m+n\right) !}{2^{m+n+1}m!n!};
\label{TTT} \\
\int \limits_{0}^{\infty }udu\exp \left( -2u\right) L_{n-1}^{1}\left(
u\right) L_{m}^{1}\left( u\right) &=&\frac{\left( m+n\right) !}{%
2^{m+n+1}m!\left( n-1\right) !}\text{,}  \notag
\end{eqnarray}%
the gap equation takes a form,%
\begin{equation}
\frac{1}{\lambda }=\frac{\overline{\omega }_{c}^{2}}{4\overline{\mu }}\sum
\nolimits_{s}\left \{
\begin{array}{c}
\sum_{n,m=0}\frac{\left( m+n\right) !}{2^{m+n}m!n!}\left( \frac{\overline{%
\omega }_{s}^{2}+\overline{\mu }^{2}}{\left( \left( -i\overline{\omega _{s}}+%
\overline{\mu }\right) ^{2}-\overline{\omega }_{c}^{2}\left( n+1\right)
\right) \left( \left( i\overline{\omega }_{s}+\mu \right) ^{2}-\overline{%
\omega }_{c}^{2}\left( 1+n\right) \right) }+\frac{\overline{\omega }_{s}^{2}+%
\overline{\mu }^{2}}{\left( \left( -i\overline{\omega }_{s}+\overline{\mu }%
\right) ^{2}-\overline{\omega }_{c}^{2}n\right) \left( \left( i\overline{%
\omega }_{s}+\overline{\mu }\right) ^{2}-\overline{\omega }_{c}^{2}m\
\right) }\right) + \\
+\sum_{n,m=1}\frac{\left( m+n\right) !}{2^{m+n}m!n!}\left( \frac{n\overline{%
\omega }_{c}^{2}}{\left( \left( -i\overline{\omega }_{s}+\overline{\mu }%
\right) ^{2}-\overline{\omega }_{s}^{2}\left( n+1\right) \right) \left(
\left( i\overline{\omega }_{s}+\overline{\mu }\right) ^{2}-\overline{\omega }%
_{c}^{2}m\right) }+\frac{m\overline{\omega }_{c}^{2}}{\left( \left( -i%
\overline{\omega }_{s}+\overline{\mu }\right) ^{2}-\overline{\omega }%
_{c}^{2}n\right) \left( \left( i\overline{\omega }_{s}+\overline{\mu }%
\right) ^{2}-\overline{\omega }_{c}^{2}\left( 1+m\right) \right) }\right)%
\end{array}%
\right \} ,  \label{GGG}
\end{equation}%
where the effective dimensionless electron - electron coupling $\lambda
=g^{2}\mu /4\pi v^{2}$. It is also convenient to scale $\mu $ and $\omega
_{c}$ by the temperature, $\overline{\mu }=\mu /T,\overline{\omega }%
_{c}=\omega _{c}/T$. After summation over the Matsubara frequency, one
obtains, separating the zero LL ($n=0$) from the rest,
\begin{equation}
\frac{1}{\lambda }=\frac{\overline{\omega }_{c}^{2}}{4\overline{\mu }}\left
\{ \sum \limits_{n,m}\frac{\left( m+n\right) !}{2^{m+n+1}}\frac{f\left[ n%
\right] f\left[ m\right] }{m!n!}s_{nm}+\sum \limits_{n}\frac{f\left[ n\right]
f\left[ 0\right] }{2^{n}}s_{n}+\frac{f\left[ 0\right] ^{2}}{2}s\right \}
\text{,}  \label{1/lamda}
\end{equation}%
where $f\left( n\right)$ will be discussed in the next subsection. The
separation is required since the expressions in Appendix B are ambiguous for
$n=0$ and should be defined using L'Hopital's rule. The $n,m>0$ part (free
of the "ambiguous" terms) is:%
\begin{eqnarray}
s_{nm} &=&A\left[ \overline{\omega }_{c}^{2}\left( n+1\right) ,\overline{%
\omega }_{c}^{2}\left( m+1\right) \right] +A\left[ \overline{\omega }%
_{c}^{2}n,\overline{\omega }_{c}^{2}m\right] +  \label{Umn} \\
&&+\left(
\begin{array}{c}
\overline{\mu }^{2}B\left[ \overline{\omega }_{c}^{2}\left( n+1\right) ,%
\overline{\omega }_{c}^{2}\left( m+1\right) \right] +\overline{\mu }^{2}B%
\left[ \overline{\omega }_{c}^{2}n,\overline{\omega }_{c}^{2}m\right] \\
+n\overline{\omega }_{c}^{2}B\left[ \overline{\omega }_{c}^{2}\left(
n+1\right) ,\overline{\omega }_{c}^{2}m\  \right] +m\overline{\omega }%
_{c}^{2}B\left[ \overline{\omega }_{c}^{2}n,\overline{\omega }_{c}^{2}\left(
m+1\right) \right]%
\end{array}%
\right) \text{.}  \notag
\end{eqnarray}%
The mixed zero-nonzero LL ($n=0,m>0$) part is
\begin{equation}
s_{n}=A\left[ \overline{\omega }_{c}^{2}\left( n+1\right) ,\overline{\omega }%
_{c}^{2}\right] +A\left[ \overline{\omega }_{c}^{2}n,0\right] +\overline{\mu
}^{2}B\left[ \overline{\omega }_{c}^{2}\left( n+1\right) ,\overline{\omega }%
_{c}^{2}\right] +\overline{\mu }^{2}B\left[ \overline{\omega }_{c}^{2}n,0%
\right] \text{,}  \label{zero}
\end{equation}%
while the purely zero LL contribution
\begin{equation}
s=A\left[ \overline{\omega }_{c}^{2},\overline{\omega }_{c}^{2}\right] +A%
\left[ 0,0\right] +\overline{\mu }^{2}B\left[ \overline{\omega }_{c}^{2},%
\overline{\omega }_{c}^{2}\right] +\overline{\mu }^{2}B\left[ 0,0\right] \,%
\text{.}  \label{U}
\end{equation}%
Explicit form of functions $A$ and $B$ is given in Appendix B. It is shown
there that the functions are finite for any value of magnetic field and
temperature $T>0$. The sum is computed numerically.

\subsection{Phonon retardation effects}

Usually within the BCS approach, the interaction is approximated not just by
a contact in space and a step function - like cutoff,%
\begin{equation}
\mu -\hbar \Omega <\hbar \omega _{c}\sqrt{n}<\mu +\hbar \Omega \text{,}
\label{shell}
\end{equation}%
see Fig.1. Therefore the sums over Landau levels in Eq.(\ref{1/lamda}) is
restricted. The approximation is not good enough for our purposes, since,
when crossing a Landau level by increasing the field infinitesimally, the
result of summation in the quantum regime jumps by a finite amount like Hall
conductivity in 2DEG. This is unphysical since the step function dependence
is just an approximation of a more realistic second order effective electron
interaction due to phonon exchange.

Neglecting the dispersion of the optical phonon, the sharp cutoff will be
replaced by the Lorentzian function of $\omega _{s}=\pi T\left( 2s+1\right)
/\hbar $:
\begin{equation}
V\left( s,p\right) =\frac{g^{2}\Omega ^{2}}{\Omega ^{2}+\omega _{s}^{2}}.
\label{Lorentzian}
\end{equation}%
In our scaled units the summation over Landau levels comes with a weight
function,%
\begin{equation}
f\left( n\right) =\frac{\Omega ^{2}}{\Omega ^{2}+\left( \omega _{c}\sqrt{n}%
-\mu /\hbar \right) ^{2}}\text{.}  \label{fn}
\end{equation}%
The remaining sums over Landau levels in Eq.(\ref{Umn}) were performed
numerically to determine the normal - superconductor transition line.

\begin{figure}[tbp]
\begin{center}
\includegraphics[width=12cm]{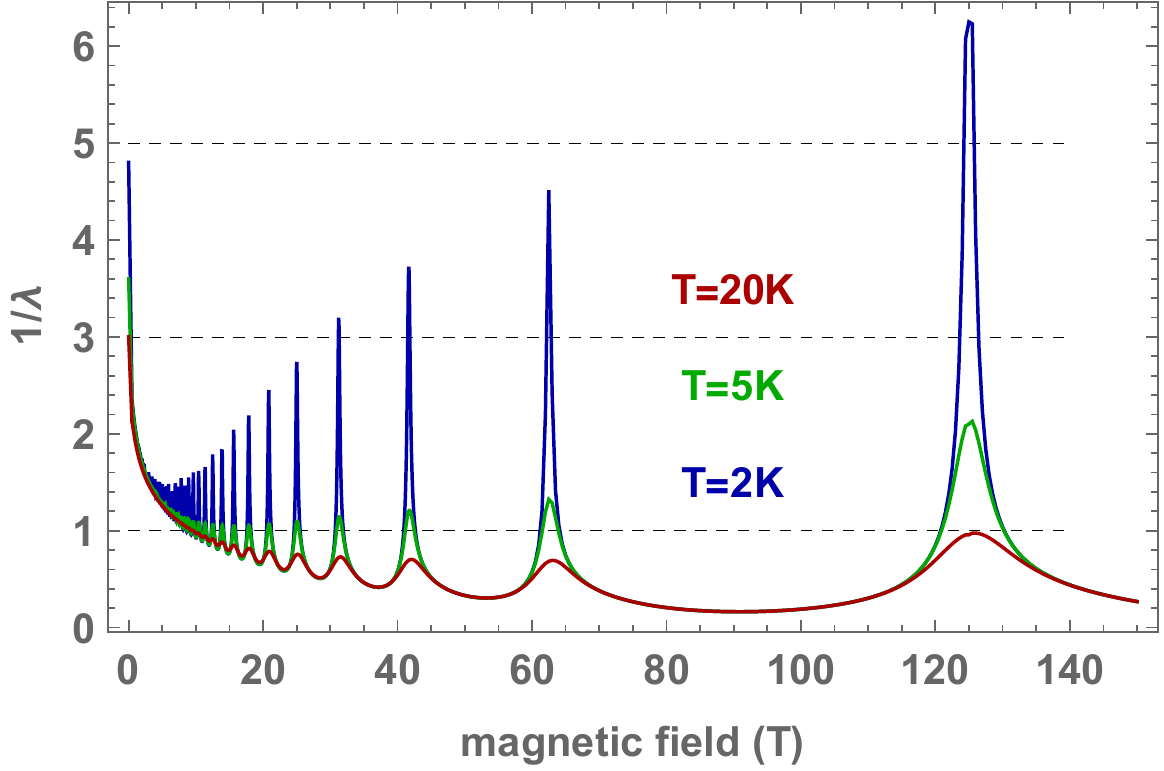}
\end{center}
\par
\vspace{-0.5cm}
\caption{The inverse effective electron coupling is presented for three
temperatures $\hbar \Omega /200$, $\hbar \Omega /50$,$\hbar \Omega /20$, in
a wide range of magnetic field up to $15\hbar c\Omega ^{2}/ev^{2}$. The
value of chemical potential is chosen as at $\protect \mu =5\hbar \Omega $. }
\end{figure}
\begin{figure}[tbp]
\begin{center}
\includegraphics[width=16cm]{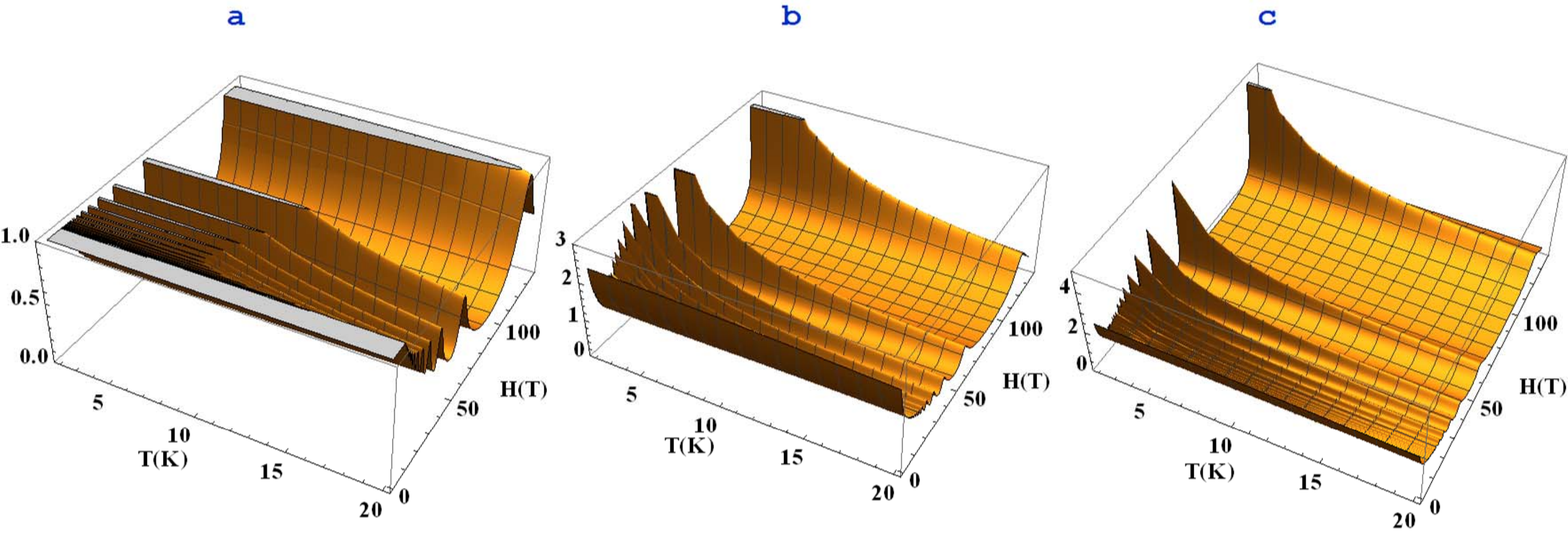}
\end{center}
\par
\vspace{-0.5cm}
\caption{The fragmented $H-T$ phase diagram of the 2D Weyl semi - metal.
Cross-sections (in gray) outline the superconducting "domed". Three values
of the effective electron - electron coupling are given. a. $\protect \lambda %
=1.$ b. $\protect \lambda =0.33.$ c. $\protect \lambda =0.2$.}
\end{figure}

\subsection{The fragmented transition line}

Magnetic phase diagram is the main result of the present paper. Although in
experiments the material parameter $\lambda $ is fixed, while temperature
and magnetic field (or both) are external parameters, it is more convenient
to calculate the critical value of $\lambda $ as a function of temperature
and magnetic field. In Fig. 2 the inverse effective electron - electron
coupling $\lambda ^{-1}$ is plotted as a function of magnetic field. Curves
correspond to three temperatures $\hbar \Omega /200$, $\hbar \Omega /50$, $%
\hbar \Omega /20$, while the wide range of magnetic fields extends up to $%
25\hbar c\Omega ^{2}/ev^{2}$. The value of chemical potential is chosen to
be $\mu =5\hbar \Omega $. To concreteness (and to facilitate a discussion of
an experiment on $Cd_{3}As_{2}$) we use typical values of the Debye
frequency $\Omega =400K$ and the Fermi velocity $v=10^{8}cm/s$,$\ $so that
temperatures and fields in Fig. 2 are given in kelvins and tesla
respectively. Dashed lines mark the cases of a weak, $\lambda =0.2$, an
intermediate, $\lambda =0.33,$and a relatively strong coupling $\lambda =1$.

For the weak coupling the conventional $H_{c2}$ does not appear in the
figure, since the critical temperature is below $2K$. The only
superconducting "dome" appears at the quantum limit with Cooper pairs made
on the lowest LL only. At the intermediate coupling the conventional $%
H_{c2}=2T$ does appear (around $4K$), but now there are four additional
superconducting domes at Landau levels $N=1-4$. At the strong coupling
regular $H_{c2}$ around $12T$ is clearly the dominant feature with numerous
domes appearing at $T=2K$. The problematic issue of rigorously defining the
semi-classical notion of $H_{c2}$ from the microscopic calculation is the
same as for the conventional superconductor (parabolic band)\cite{Indians}.
Of course at yet lower temperatures more domes appear.

In Fig.3 the phase diagram in the $H-T$ is presented for the same three
values of the effective electron - electron couplings.

The superconducting domes on Landau levels are clearly seem as gray areas.
Generally they become very narrow as the LL index $N$ grows, at low
temperatures and at weak couplings. The WSM, in which we suspect that the
high magnetic field superconducting domes were observed (see Section IV),
are anisotropic 3D WSM. In addition at fields as large as $50-60T$ applied
in recent experiments\cite{Cao,Wang} the Zeeman coupling to spin cannot be
ignored. Therefore the next section is devoted to generalizations to the
direct coupling to the electron spin and to 3D WSM.

\section{Generalizations: Zeeman coupling and 3D WSM.}

\subsection{Zeeman coupling, the paramagnetic limit}

Along with the orbital effect of magnetic field on electrons and their
pairing, at very high fields the direct (Zeeman) coupling of the magnetic
field to spin becomes significant. A textbook example is the Chandrasekhar -
Klogston\cite{Chandrasekhar} pair breaking phenomenon in conventional
metallic (parabolic single band) superconductors.

To investigate the Zeeman coupling effect on superconductivity in (2D) WSM,
let us consider the following Hamiltonian%
\begin{equation}
H=K+K_{Z}+V\text{.}  \label{H_Z}
\end{equation}%
Here the kinetic energy term and the phonon mediated effective interaction
are still defined in Eq.(\ref{K}) and Eq.(\ref{int}) respectively. The
Zeeman coupling term is

\begin{equation}
\text{\  \  \ }K_{Z}=-g_{L}\mu _{B}H\int_{\mathbf{r}}\psi _{\alpha }^{s\dagger
}\left( \mathbf{r}\right) \tau _{st}^{z}\delta _{\alpha \beta }\psi _{\beta
}^{t}\left( \mathbf{r}\right) \text{,}  \label{K_Z}
\end{equation}%
where $\tau _{st}^{z}$ is the Pauli matrix in spin space, $g_{L}$ and $\mu
_{B}$ are the Lande factor and the Bohr magneton respectively.

A simple singlet Ansatz, Eq.(\ref{spin Ansatz}), no longer solves the set of
the Gor'kov equations. Therefore they should be explicitly solved. The
number of the Greens functions in this case is doubled compared to the case
considered in Section III. However the phase Ansatz for GF in magnetic
field, Eq.(\ref{n3a}), still holds. Substituting Eq.(\ref{n3a}) into the gap
equation (see Eq.(\ref{C3}) of Appendix C, where derivations also can be
found), and using a pseudospin singlet Ansatz for the gap function, $\Delta
_{\alpha \gamma }^{\ast }\left( \mathbf{r}\right) =\sigma _{\alpha \gamma
}^{x}\exp \left( -r^{2}/2l^{2}\right) $, one obtains equation for critical
curve in $H-T$ plane:

\begin{equation}
\frac{1}{\pi g^{2}}=T\sum \nolimits_{\omega }\int_{\rho }\rho e^{-\frac{\rho
^{2}}{2l^{2}}}\left(
\begin{array}{c}
\ g_{21}^{2\downarrow \downarrow }\left( \mathbf{-}\rho \right) \
g_{21}^{1\uparrow \uparrow }\left( \rho \right) +\ g_{22}^{2\uparrow
\uparrow }\left( \mathbf{-}\rho \right) \ g_{11}^{1\downarrow \downarrow
}\left( \rho \right) +\ g_{22}^{2\downarrow \downarrow }\left( \mathbf{-}%
\rho \right) \ g_{11}^{1\uparrow \uparrow }\left( \rho \right)
+g_{21}^{2\uparrow \uparrow }\left( \mathbf{-}\rho \right) \
g_{21}^{1\downarrow \downarrow }\left( \rho \right) \\
\ g_{11}^{2\downarrow \downarrow }\left( \mathbf{-}\rho \right) \
g_{22}^{1\uparrow \uparrow }\left( \rho \right) +\ g_{11}^{2\uparrow
\uparrow }\left( \mathbf{-}\rho \right) \ g_{22}^{1\downarrow \downarrow
}\left( \rho \right) +\ g_{12}^{2\downarrow \downarrow }\left( \mathbf{-}%
\mathbb{\rho }\right) \ g_{12}^{1\uparrow \uparrow }\left( \rho \right) +\
g_{12}^{2\uparrow \uparrow }\left( \mathbf{-}\rho \right) \
g_{12}^{1\downarrow \downarrow }\left( \rho \right)%
\end{array}%
\right) \text{.}  \label{Spara}
\end{equation}%
The set of the spin dependent GF is calculated in Appendix C (Eqs.(\ref{C7}%
),(\ref{C8})).

Substituting them into Eq.(\ref{Spara}) performing integration over $\rho $,
and summation on Matsubara frequencies one obtain relation for critical
curve at the $\lambda ^{-1}-H$ plane:
\begin{equation}
\frac{1}{\lambda }=\frac{\overline{\omega }_{c}^{2}}{4\overline{\mu }}%
\sum_{n=0,m=0}^{\infty }\frac{\left( m+n\right) !}{2^{m+n}}\frac{f\left[ n%
\right] f\left[ m\right] }{m!n!}s_{nm}\text{.}  \label{paracurve}
\end{equation}%
Here functions $s_{nm}$ are,
\begin{eqnarray}
s_{nm} &=&A_{p}\left[ \overline{\omega }_{c}^{2}\left( n+1\right) ,\overline{%
\omega }_{c}^{2}\left( m+1\right) ,\overline{\mu }+\overline{\varepsilon },%
\overline{\mu }-\overline{\varepsilon }\right] +A_{p}\left[ \overline{\omega
}_{c}^{2}\left( n+1\right) ,\overline{\omega }_{c}^{2}\left( m+1\right) ,%
\overline{\mu }-\overline{\varepsilon },\overline{\mu }+\overline{%
\varepsilon }\right] +  \label{smn} \\
&&+A_{p}\left[ \overline{\omega }_{c}^{2}n,\overline{\omega }_{c}^{2}m,%
\overline{\mu }+\overline{\varepsilon },\overline{\mu }-\overline{%
\varepsilon }\right] +A_{p}\left[ \overline{\omega }_{c}^{2}n,\overline{%
\omega }_{c}^{2}m,\overline{\mu }-\overline{\varepsilon },\overline{\mu }+%
\overline{\varepsilon }\right] +  \notag \\
&&+\left( \overline{\mu }^{2}-\overline{\varepsilon }^{2}\right) \left(
\begin{array}{c}
B_{p}\left[ \overline{\omega }_{c}^{2}\left( n+1\right) ,\overline{\omega }%
_{c}^{2}\left( m+1\right) ,\overline{\mu }+\overline{\varepsilon },\overline{%
\mu }-\overline{\varepsilon }\right] \\
+B_{p}\left[ \overline{\omega }_{c}^{2}\left( n+1\right) ,\overline{\omega }%
_{c}^{2}\left( m+1\right) ,\overline{\mu }-\overline{\varepsilon },\overline{%
\mu }+\overline{\varepsilon }\right] \\
+B_{p}\left[ \overline{\omega }_{c}^{2}n,\overline{\omega }_{c}^{2}m,%
\overline{\mu }+\overline{\varepsilon },\overline{\mu }-\overline{%
\varepsilon }\right] +B_{p}\left[ \overline{\omega }_{c}^{2}n,\overline{%
\omega }_{c}^{2}m,\overline{\mu }-\overline{\varepsilon },\overline{\mu }+%
\overline{\varepsilon }\right]%
\end{array}%
\right)  \notag \\
&&+n\overline{\omega }_{c}^{2}\left[ B_{p}\left[ \overline{\omega }%
_{c}^{2}\left( n+1\right) ,\overline{\omega }_{c}^{2}m,\overline{\mu }+%
\overline{\varepsilon },\overline{\mu }-\overline{\varepsilon }\right]
+nB_{p}\left[ \overline{\omega }_{c}^{2}\left( n+1\right) ,\overline{\omega }%
_{c}^{2}m,\overline{\mu }-\overline{\varepsilon },\overline{\mu }+\overline{%
\varepsilon }\right] \right]  \notag \\
&&+mB_{p}\left[ \overline{\omega }_{c}^{2}n,\overline{\omega }_{c}^{2}\left(
m+1\right) ,\overline{\mu }+\overline{\varepsilon },\overline{\mu }-%
\overline{\varepsilon }\right] +mB_{p}\left[ \overline{\omega }_{c}^{2}n,%
\overline{\omega }_{c}^{2}\left( m+1\right) ,\overline{\mu }-\overline{%
\varepsilon },\overline{\mu }+\overline{\varepsilon }\right] \text{,}  \notag
\end{eqnarray}%
where the dimensionless ratio of the Zeeman energy and temperature, $%
\overline{\varepsilon }=2g_{L}\mu _{B}H/T$, is used. In the spin
non-degenerate case the separation of the zero LL is not required due to the
difference in chemical potentials of the spin projections. The Matsubara
sums read:

\begin{eqnarray}
A_{p}\left[ a,b,\mu _{1},\mu _{2}\right] =\frac{1}{4\sqrt{a}}\left \{ \frac{%
\left( \sqrt{a}-\mu _{1}\right) ^{2}\tanh \left[ \frac{\sqrt{a}-\mu _{1}}{2}%
\right] }{\left( \sqrt{a}-\mu _{1}-\mu _{2}\right) ^{2}-b}+\frac{\left(
\sqrt{a}+\mu _{1}\right) ^{2}\tanh \left[ \frac{\sqrt{a}+\mu _{1}}{2}\right]
}{\left( \sqrt{a}+\mu _{1}+\mu _{2}\right) ^{2}-b}\right \}  &&+\left(
\begin{array}{c}
a\longleftrightarrow b \\
\mu _{1}\longleftrightarrow \mu _{2}%
\end{array}%
\right) ; \\
B_{p}\left[ a,b,\mu _{1},\mu _{2}\right] =\frac{1}{4\sqrt{a}}\left \{ \frac{%
\tanh \left( \frac{\sqrt{a}-\mu _{1}}{2}\right) }{\left( \sqrt{a}-\mu
_{1}-\mu _{2}\right) ^{2}-b}+\frac{\tanh \left( \frac{\sqrt{a}+\mu _{1}}{2}%
\right) }{\left( \sqrt{a}+\mu _{1}+\mu _{2}\right) ^{2}-b}\right \}
&&-\left(
\begin{array}{c}
a\longleftrightarrow b \\
\mu _{1}\longleftrightarrow \mu _{2}%
\end{array}%
\right) \text{.}  \notag
\end{eqnarray}

The results of numerical calculations are presented in Fig. 4. The inverse
effective coupling $\lambda ^{-1}$ as function of magnetic field for six
values of the material parameter characterizing the strength of the Zeeman
coupling on superconductivity, $\alpha _{p}=g_{L}\mu _{B}c\Omega /ev^{2}$, $%
\alpha _{p}=2\cdot 10^{-4},5\cdot 10^{-4},1.5\cdot 10^{-3},3.5\cdot
10^{-3},3.5\cdot 10^{-3},1.7\cdot 10^{-2}$, are plotted. Temperature is
fixed at $T=0.005\hbar \Omega $ (as above, we take $\hbar \Omega =400K$ for
concretions this amounts to $2K$), $\mu =5\hbar \Omega $, while the range of
magnetic fields is between $5\hbar c\Omega ^{2}/ev^{2}$ to $30\hbar c\Omega
^{2}/ev^{2}$. For a typical value of the Fermi velocity $c=10^{8}cm/s$ this
corresponds to $25-150T$. The magnetic phase\ ($H-T$) diagram is obtained,
as in the previous section, as a set of fields for a fixed $\lambda $.

\begin{figure}[tbp]
\begin{center}
\includegraphics[width=16cm]{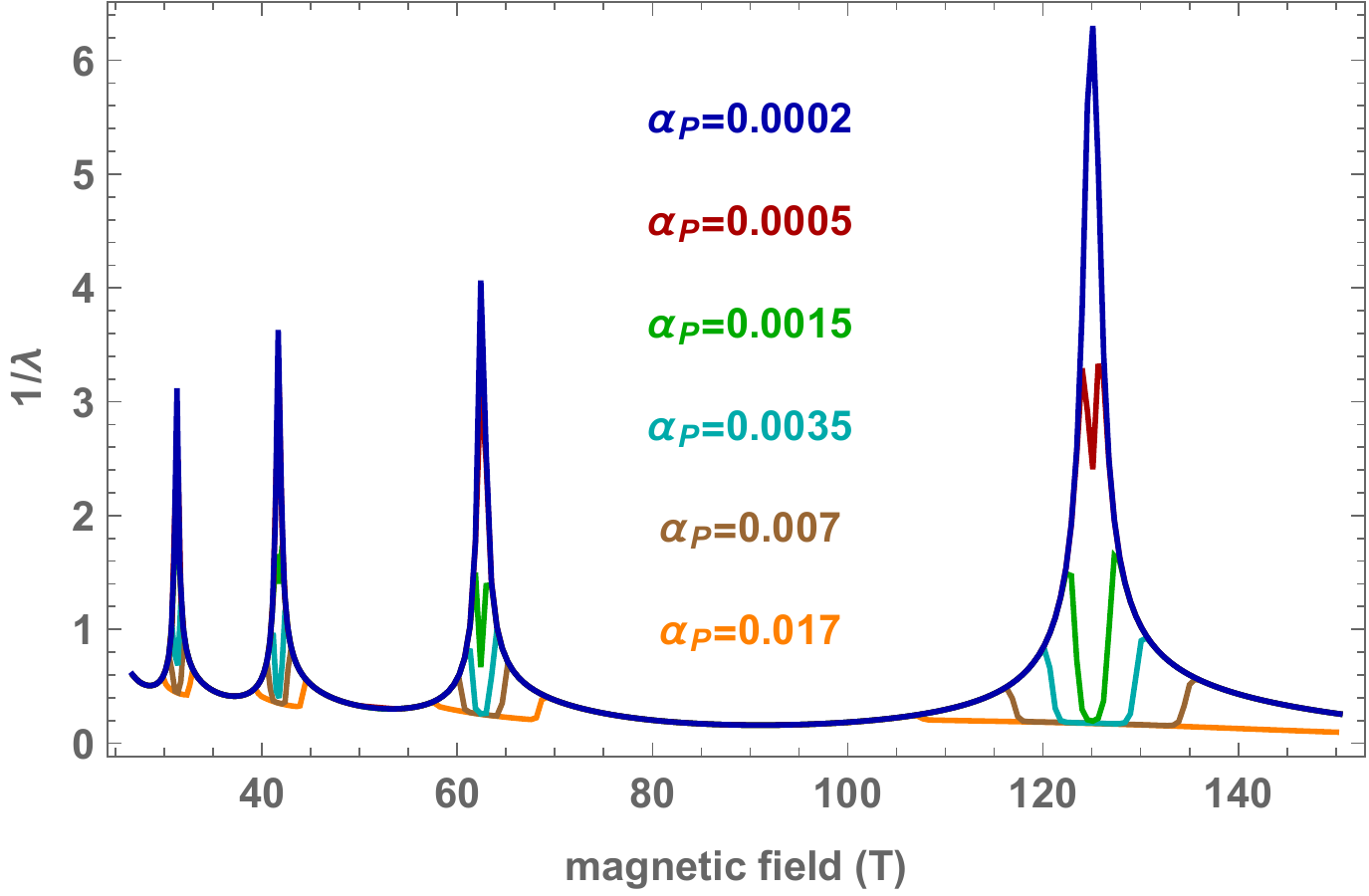}
\end{center}
\par
\vspace{-0.5cm}
\caption{Superconductor - normal critical curve in the $\protect \lambda %
^{-1}-H\ $plane. Zeeman interaction splits the superconducting domes
suppressing superconductivity at large values of the dimensionless of the
paramagnetic coefficient $\protect \alpha _{p}=g_{L}\protect \mu _{B}c\Omega
/ev^{2}$.}
\end{figure}

One observes that while for the smallest Zeeman coupling (blue curve) there
is no difference with the zero Zeeman splitting case (blue line in Fig.2),
for the largest value the superconductivity is quenched due Chandrasekhar-
Klongstone (paramagnetic) limit. For the intermediate values of $\alpha _{p}$
splitting of the superconducting domes of the fractured critical line is
well pronounced.

Band structure calculations of one of the most promising WSM $Cd_{3}As_{2}$
show\cite{Jeon} that the Dirac point in this system is formed by the spin
mixed with the sublattice index. In this case the Zeeman interaction with
the external magnetic field is more complicated than considered in our two
band model. It is reasonable to expect however that qualitative features of
the Zeeman coupling are similar. Another important characteristics of
superconducting WSM is that many of them are three dimensional.

\subsection{Generalization to 3D WSM}

In this subsection the calculation of the magnetic phase diagram is
generalized to 3D WSM with (typically several) Dirac points. The band
structure of an asymmetric 3D WSM near such a point is captured by the
Hamiltonian
\begin{equation}
K=\int_{\mathbf{r}}\psi _{\alpha }^{s\dagger }\left( \mathbf{r,}z\right)
\left \{ -i\hbar v\left( D_{x}\sigma _{\alpha \beta }^{x}+D_{y}\sigma
_{\alpha \beta }^{y}\right) -i\hbar v_{z}\partial _{z}\sigma _{\alpha \beta
}^{z}-\mu \delta _{\alpha \beta }\right \} \psi _{\beta }^{s}\left( \mathbf{%
r,}z\right) \text{.\  \  \  \ }  \label{K_3D}
\end{equation}%
Here $v$ is Fermi velocity (assumed isotropic) in the $x-y$ plane
perpendicular to magnetic field and $v_{z}$ the Fermi velocity along the
field and the gauge in the covariant derivatives is chosen to be $\mathbf{A}%
=H\left( -y/2,x/2,0\right) $. The momentum $p_{z}$ in this gauge is a
conserved quantum number.

The calculation is analogous to the 2D one, since magnetic field enters the
dependence Greens functions on lateral dimensions only. The Fourier
transform is defined now by
\begin{equation}
G_{\gamma \kappa }\left( \mathbf{r},z,\tau \right) =T\sum \nolimits_{s}\exp %
\left[ -i\omega _{s}\tau +ip_{z}z\right] G_{\gamma \kappa }\left( \omega ,%
\mathbf{r,}p_{z}\right) \text{.}  \label{Fourier 3D}
\end{equation}
It is important to distinguish between the thin film and the "bulk" cases.
For a film of thickness $d$, the field component of the "momentum" is
discretized as:

\begin{equation}
p_{z}=\frac{\pi \hbar }{d}M,M=\pm 1,2...  \label{pz}
\end{equation}%
The equations for two normal GF (see Eqs.(\ref{EqH})) in the 3D case read:

\begin{eqnarray}
\left[ iv\mathbf{D}_{r}^{i}\cdot \mathbf{\sigma }_{\gamma \beta
}^{i}-v_{z}p_{z}\sigma _{\gamma \beta }^{z}+\left( i\omega +\mu \right)
\delta _{\gamma \beta }\right] G_{\beta \kappa }^{1}\left( \mathbf{r,r}%
^{\prime },p_{z}\right) &=&\delta ^{\gamma \kappa }\delta \left( \mathbf{r-r}%
^{\prime }\right) \text{,}  \label{3D_Gorkov} \\
\left[ -iv\mathbf{D}_{r}^{i}\cdot \mathbf{\sigma }_{\gamma \beta
}^{ti}+v_{z}p_{z}\sigma _{\gamma \beta }^{z}+\left( -i\omega +\mu \right)
\delta _{\gamma \beta }\right] G_{\beta \kappa }^{2}\left( \mathbf{r,r}%
^{\prime },p_{z}\right) &=&\delta ^{\gamma \kappa }\delta \left( \mathbf{r-r}%
^{\prime }\right) \text{.}  \notag
\end{eqnarray}%
The magnetic phase Ansatz Eq.(\ref{n3a}) still solves the 3D gap equation
Eq.(\ref{EqH}) (see Appendix D). Moreover the gaussian form of the gap
function (independent of $z$), Eq.(\ref{deltaAnsatz}) is not changed. The
equation determining the critical curve in the $H-T$ plane is now:

\begin{equation}
\frac{1}{\lambda }=\frac{\zeta \overline{\omega }_{c}^{2}}{4\mu ^{2}}\sum
\limits_{M>0}\left \{ \sum \limits_{n,m}\frac{\left( m+n\right) !}{2^{m+n+1}}%
\frac{f\left[ n\right] f\left[ m\right] }{m!n!}s_{nmM}+\sum \limits_{n}\frac{%
f\left[ n\right] f\left[ 0\right] }{2^{n}}s_{nM}+\frac{f\left[ 0\right] ^{2}%
}{2}s_{M}\right \} \text{.}  \label{3D domes}
\end{equation}%
Here the 3D effective attraction strength (see Appendix D for the relevant
DOS) is $\lambda =g^{2}\mu ^{2}/2\pi ^{2}v_{z}v^{2}$ and the dimensionless
parameter inversely proportional to the thickness is defined by $\zeta =\pi
v_{z}/dT$. The functions $s_{nmM},s_{nM},s_{M}$ depending on the new quantum
number $M$, defined in Eq.(\ref{pz}) and details of derivation (including
the relevant GF in this case) are given in Appendix D, while the function $f$
containing the frequency dependence of the effective phonon mediated
interaction remains as in 2D, see Eq.(\ref{fn}).

The result for films of two values of the film thickness corresponding to
values $\zeta =0.021$ and $\zeta =0.11$ and fixed temperature $T=0.005\hbar
\Omega $ (for $\Omega =400K$ it amount to $T=2K$) are presented in Fig. 5a
and 5b respectively. They demonstrate essential transformation of the
superconducting - normal fractured critical line compared to the 2D case.
The smaller value of $\zeta $ practically corresponds to the bulk, while the
larger represents a thin film. In the bulk the domes become asymmetric due
to the dispersion along the field. Generally larger coupling $\lambda $ is
required to create the superconducting state on the Landau levels. The
phenomenon of the re - entrant superconductivity itself however is clearly
present due to enhancement of the DOS despite the fact that in 3D DOS does
not vanishes between the LL.

\begin{figure}[tbp]
\begin{center}
\includegraphics[width=16cm]{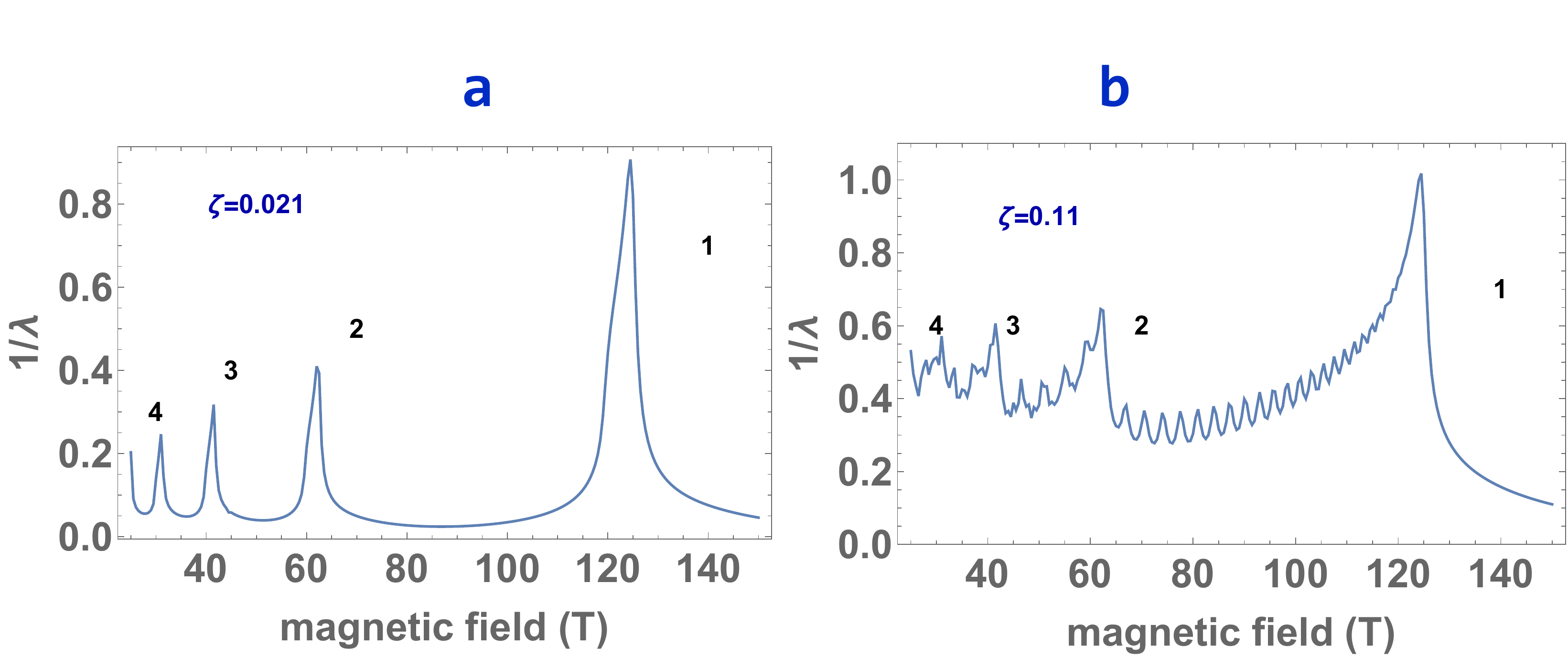}
\end{center}
\par
\vspace{-0.5cm}
\caption{Critical curve in the inverse coupling - magnetic field ($\protect%
\lambda ^{-1}-H$) plane at fixed temperature for 3D WSM. The temperature
value is $T=0.005\  \hbar \Omega $. a. Thick slab, $\protect \zeta \equiv
\protect \pi \hbar v_{z}/dT=0.021$. b. Thin film, $\protect \zeta =0.11$. $%
v_{z}$ is the electron velocity in magnetic field direction.}
\end{figure}

The superconducting "domes" become wider in slab geometry (Fig. 5a) and
demonstrate in set of small secondary peaks (ripples) caused by the
quantization of the momentum along the field ($p_{z}$) direction in a thin
film. Higher LL disappear. To conclude in the bulk the third dimension
"smooths" the effect on Landau quantization as it appears in 2D, but just
slightly, while in thin films the shape is modified.

\section{Comparison with experiments, discussion and conclusions}

In this section experimental evidence for existence of the Cooper pairing in
WSM above $H_{c2}$ is discussed. In addition we discuss the various tacit
assumptions of our model and theoretical methods: speculate on possible
transition to a triplet superconducting phase and a necessity to go beyond
the adiabatic approximation used in the present paper. The conventional
metals are explicitly contrasted with Weyl semi - metals.

\subsection{Magnetoresistance as a signature of the superconducting state at
Landau levels}

A "smoking gun" revealing the existence of superconductivity on Landau
levels would be the dependence of resistivity on magnetic field. In normal
metal one observes the resistivity generally increase faster than $H$
superimposed with Shubnikov deHaas (SdH) oscillations around Landau levels.
The picture is supported by detailed semi-classical theory valid for high
Landau levels\cite{Abrikosov2}. In the present paper the superconductivity
in the quantum limit was studied. How will it influence the
magnetoresistance at previously unreachable fields of order 100T beyond the
semiclassical regime?

Inside the "superconducting "domes" (constituting a very tiny fraction of
the magnetic phase diagram within the narrow range of fields)
magnetoresistance does not vanish due to phenomenon of the "flux flow".
Since 3D Weyl semi-metals can be made very clean, unpinned vortex liquid
rather than pinned vortex glass \cite{vortexglass} is formed. When vortices
are allowed to move, the dissipation inside the cores ensues, but the flux
flow resistivity is much smaller than the normal state. In the vortex glass
state the effect would be more dramatic: the resistivity drops (almost) to
zero. It should be noted that "vortices" in the present context should be
understood as an inhomogeneity of the order parameter, since the magnetic
"envelop" (of the size of magnetic penetration depth) of multiple vortices
strongly overlap at such fields. As a result magnetization is practically
homogeneous\cite{RMP,Maniv}. Damping of the amplitude of SdH oscillations in
superconducting regions is not expected to be significant, as was already
noted while analyzing the SdH oscillations in organic superconductor\cite%
{Wosnitza} below the upper critical field of $3.6T$. The physics of the
superconducting state on the LL is still insufficiently studied (only the
quantum limit for the parabolic band material was theoretically described in
a series works\cite{Ducan})

In a remarkable experiments \cite{Cao} with magnetic fields up to $50T$ it
was found that beyond several SdH oscillations at high LL riding on
magnetoresistance quadratic in $H$ ($N=6-15$ are clearly seen at $T=3K$),
upon approaching quantum limit at $N=2-4$ the magnetoresistance levels off.
The amplitude of the oscillations gradually increases. It is very difficult
to explain why the fast increase of the magnetoresistivity is halted at $%
10-20$ $T$. It is natural to interpret this as appearance of
superconductivity as in Fig. 2 for moderate $\lambda $. Indeed the
superconductivity (in the dynamic vortex liquid flux flow phase) would
strongly reduce the magnetoresistance magnitude. Our calculation is 2D,
however the effect of 3D in strong magnetic field is rather minor: the peaks
in Fig. 2,3 will be broadened. In the experiment at $N=2,3$ a significant
Zeeman and pseudospin splitting (with and accompanying Berry phase) are
observed and these will be discussed below. The splitting is clearly seen in
magnetoresistance data of ref.\onlinecite{Cao} at fields above $25T$.

Similar phenomenon (less pronounced since applied magnetic fields were up to
$16T$ only) was observed\cite{TaP} in Weyl superconductor $TaP$ above $%
H_{c2} $ (A quite conventional magnetic phase diagram was experimentally
established in this materials with $H_{c2}\left( 1K\right) =3T$ and $%
T_{c}=3.5K)$. As before, the fast increase of magnetoresistance is leveled
off at small $N$. Unfortunately it is difficult assign definite $N$ to SdH
oscillations clearly seen at $T=3K$. This would correspond to weak coupling
case shown in Fig. 2,3. The same relates to the recent discovery of
"logarithmic series" of oscillations\cite{ZrTelow} in the same material at
density of order $10^{16}$. The quantum limit is reached and leveling of
magnetoresistance is observed, but if superconductivity is formed at low
Landau levels it is nonadiabatic (see below).

\subsection{On the possibility of the triplet pairing}

Our calculation was restricted to the singlet pairing. In some cases strong
magnetic field might in principle favor triplet, however there is no
experimental evidence in 3D Weyl semi-metals for a triplet state so far. One
therefore can ask the following question: is the triplet state possible
theoretically in models of WSM considered here. The question was addressed
theoretically in a slightly different context of the 2D WSM surface state of
topological insulator\cite{Fu}\cite{JCM15}. In this system it was found that
both the singlet and the triplet phases exist. However, although they are
nearly degenerate in some cases (very small chemical potential $\mu $), the
singlet always prevail energetically. It was also shown theoretically\cite%
{Magnetic impur} that magnetic impurities or proximity to the Stoner
instability (local magnetic moment due to the exchange interaction) can
favor the triplet state. In such case the triplet superconducting state in
WSM must survive in extremely strong magnetic fields.

Another strong argument in 3D was put forward long ago by Rasolt and
Tesanovic\cite{RMP}. They argued that the Chandrasekhar - Klogston breaking
of the singlet state is ineffective due to spacial inhomogeneity of the
order parameter in the field direction. This remains valid for WSM.

\subsection{Comparison of the WSM superconductor to conventional parabolic
band superconductor, adiabatic approximation}

Let us complement the qualitative estimates made in the introduction on the
comparison between the pairing on Landau levels in the parabolic band
materials (including semi - metals\cite{Koshelev}) and WSM by contrasting
the magnetic phase diagrams. In Fig. 6 the phase diagram of the 2D single
parabolic band superconductor with the electron-phonon coupling $g$, Debye
frequency $\Omega $ and the chemical potential $\mu =5\hbar \Omega $ as for
WSM is Section III (see blue curve in Fig. 2) is presented. The effective
mass of the conventional metal is assumed to be equal to that of the free
electron mass. The inverse effective coupling $\lambda ^{-1}$ (calculated
with pertinent density of states) is given as function of magnetic field at
the same temperatures $T=0.005,0.0125,0.05\hbar \Omega $ (corresponding to $%
2,5,20K$, if $\hbar \Omega =400K$). The range of magnetic field plotted is
however much wider: $200-3000T$. The fields are necessarily super - high, if
one were to attempt the quantum limit (low Landau level) for conventional
metals, as follows from the qualitative estimate in Introduction. Inset
shows a (slightly) more accessible fields.

One observes that although in quantum limit the coupling required is not
large, field are inaccessible. On the other hand, even beyond $100T$, one
has superconducting "domes" at intermediate coupling at high LL $N>>10$ (so
that the system enters the semi - classical regime\cite{Maniv} with weak
quantization effects). The effect therefore is smeared out by disorder other
effects. Note that, as demonstrated in Fig.5, in 3D the peaks at higher LL
is further broadened and become unobservable.

Very recently superconductivity in a two parabolic band semi - metal in
strong magnetic field was considered\cite{Koshelev}. One of the bands is
quasiparticle with distance of the band edge to the Fermi level $\mu
_{e}>>\hbar \Omega $,$\ $well within the adiabatic approximation, while the
second is hole with very small $\mu _{h}<\hbar \Omega $. The Landau
quantization effect is most pronounced near the Lifshitz point, where
superconducting "domes" in magnetic phase diagram are clearly seen.

It is important to note that assumptions of our calculation include the
adiabatic pairing, namely that the Fermi level is larger than the Debye
energy $\mu /\Omega >1$. WSM like $ZrTe_{5}$ also can be tuned to small
chemical potential\cite{ZrTezero}, however to make use of the gaussian
approximation (in the BCS form or the Eliashberg form) one typically relies
on Migdal theorem\cite{Abrikosov}. Here it is questionable\cite{DasSarma}.
Therefore in the present paper only the adiabatic case $\mu /\hbar \Omega >5$
was discussed. It would be interesting to investigate what will happen
beyond this assumption since in many Dirac materials Fermi energy is very
low. For example Fermi energy in $ZrTe_{5}$ grown on\cite{ZrTezero} in
experiment in large fields up to $100T$ no oscillations were observed at
all. However in this experiment the density is below $10^{15}cm^{-3}$.

\begin{figure}[tbp]
\begin{center}
\includegraphics[width=12cm]{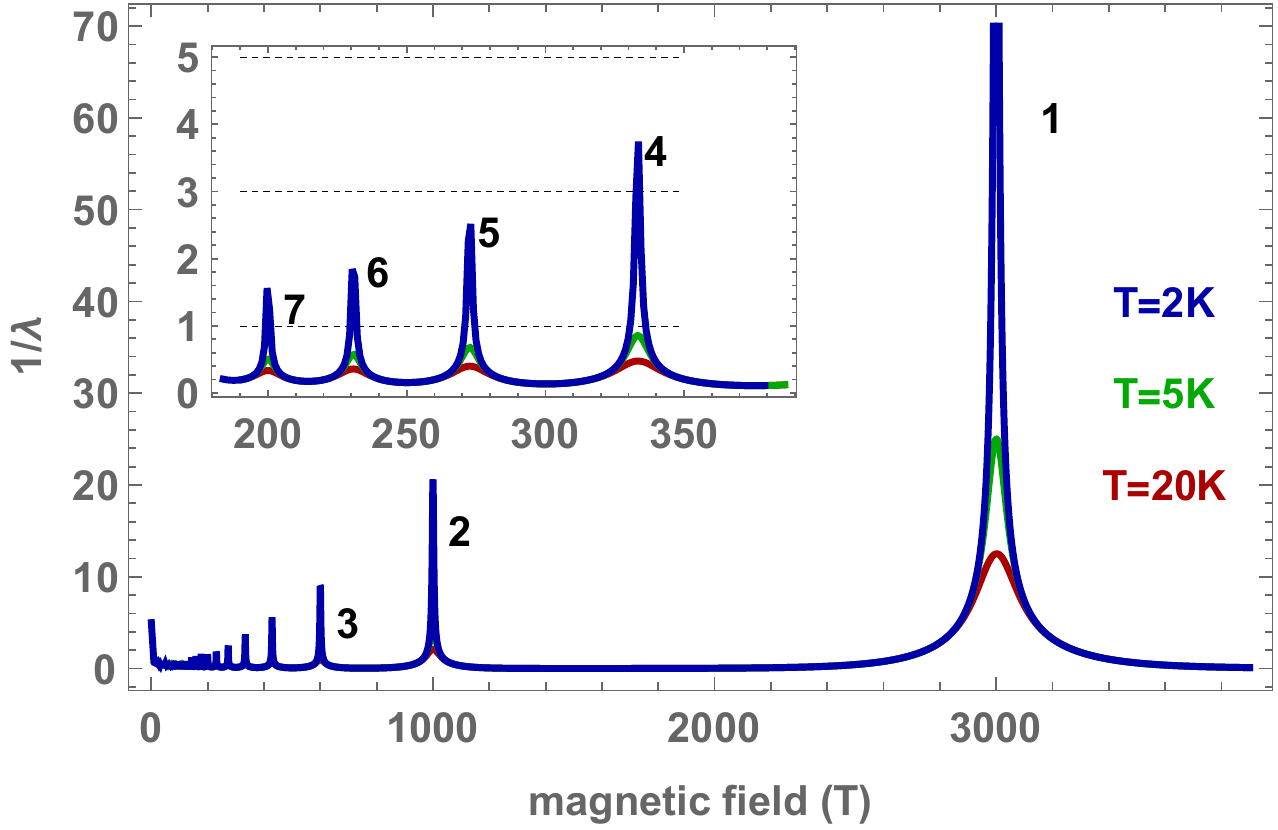}
\end{center}
\par
\vspace{-0.5cm}
\caption{Magnetic phase $\protect \lambda ^{-1}-H$ for conventional one band
metal.}
\end{figure}

\subsection{Conclusions}

Microscopic theory of phonon mediated superconductivity in Weyl semimetals
at very high magnetic fields was constructed. Weak coupling was assumed, but
the retardation effects were taken into account. It was shown that a Weyl
semi-metal in 2D and 3D that is nonsuperconducting or having a small
critical temperature $T_{c}$ at zero field becomes superconducting in narrow
regions of the magnetic phase diagram around Landau levels, especially near
the quantum limit. The Zeeman splitting sometimes becomes of significance at
highest fields. Superconductivity has an effect on magneto-conductivity
beyond conventional $H_{c2}$. Near the Landau levels the magnetoresistivity
should diminish. This might explain the recent experiments on $Cd_{3}As_{2}$
and $TaP$ and perhaps other.

This enhancement is especially pronounced for the lowest Landau level. As a
consequence, the reentrant superconducting regions in the temperature- field
phase diagram emerge at low temperatures near the magnetic fields at which
the chemical potential matches the Landau levels.

\textit{Acknowledgements.}

We are grateful to N. L. Wang, T. Maniv, T. W. Luo, V. Vinokur, J. Wang, C.
Hou, for valuable discussions.\ B.R. acknowledges MOST of ROC grant
103-2112-M-009-009-MY3 hospitality of Peking and Bar Ilan Universities. D.P.
Li was supported by National Natural Science Foundation of China (Nos.
11274018 and 11674007).

\appendix

\section{Calculation of the normal Green's functions}

In this Appendix the normal state Green's functions are calculated. In the
matrix form the equations (\ref{normalg1}), (\ref{n4ac}) read:
\begin{equation}
\widehat{h}^{a}g^{a}\left( \mathbf{\rho }\right) \ =\delta \left( \mathbf{%
\rho }\right)  \label{A3}
\end{equation}%
with 2D matrix operators $\widehat{h}^{1}=i\omega +\mu -\mathbf{\Pi }\cdot
\mathbf{\sigma ;}$ $\widehat{h}^{2}=-i\omega +\mu +\mathbf{\Pi \cdot \sigma }%
^{t}$, where $a=1,2$ and $\mathbf{\Pi =}\left \{ \mathbf{\Pi }_{x}\mathbf{%
,\Pi }_{y}\right \} $ are the ladder operators. In the symmetric gauge
\begin{equation}
\Pi _{x}=\mathbf{-}i\frac{\partial }{\partial \rho _{x}}+\  \frac{\ 1}{2l^{2}}%
\rho _{y},\Pi _{y}=-i\frac{\partial }{\partial \rho _{y}}-\frac{\ 1}{2l^{2}}%
\rho _{x}.  \label{A4}
\end{equation}%
It is convenient to rewrite them via creation and annihilation operators for
a bosonic field

\begin{equation}
a=\frac{l}{\sqrt{2}}\left( \Pi _{x}-i\Pi _{y}\right) ;a^{\dagger }=\frac{l}{%
\sqrt{2}}\left( \Pi _{x}+i\Pi _{y}\right)  \label{A5}
\end{equation}%
with the commutation relations $\left[ \Pi _{x},\Pi _{y}\right] =-i/l^{2}$, $%
\left[ a,a^{\dagger }\right] =1.$

The matrix elements of the $2\times 2\, \ $matrices $h^{a}$ are defined by
relations :%
\begin{eqnarray}
h_{11}^{1} &=&h_{22}^{1}=i\omega +\mu ;\text{ \ }h_{11}^{2}=h_{22}^{2}=-i%
\omega +\mu ;\text{\ }  \label{a6} \\
\widehat{h}_{12}^{1} &=&\widehat{h}_{21}^{2}=-\omega _{c}a;\  \  \widehat{h}%
_{21}^{1}=\widehat{h}_{12}^{2}=-\omega _{c}a^{\dagger }.  \notag
\end{eqnarray}%
Here $\omega _{c}=v\sqrt{2}/l$ is the Larmor frequency in Weyl semimetals.
Equations for normal GF can be represented in the following form
(suppressing the index $a\,$): \

\begin{eqnarray}
h_{11}g_{11}+\widehat{h}_{12}g_{21} &=&\delta \left( \mathbf{\rho }\right) ;%
\text{ }\widehat{h}_{21}g_{12}+h_{22}g_{22}=\delta \left( \mathbf{\rho }%
\right) ;  \label{A7} \\
h_{11}g_{12}+\widehat{h}_{12}g_{22} &=&0,\, \  \widehat{h}%
_{21}g_{11}+h_{22}g_{21}=0.  \notag
\end{eqnarray}%
Since $h_{11},h_{22}$ are just numbers (not operators acting on $\mathbf{%
\rho }$), one first solves the second pair of equations for the off diagonal
elements:
\begin{equation}
g_{21}=-\frac{1}{h_{22}}\widehat{h}_{21}g_{11};\text{ \ }g_{12}=-\frac{1}{%
h_{11}}\widehat{h}_{12}g_{22}\text{.}  \label{A8}
\end{equation}%
Substituting into the first pair, one obtains:
\begin{eqnarray}
\left( h_{22}h_{11}-\widehat{h}_{12}\widehat{h}_{21}\right) g_{11}\left(
\mathbf{\rho }\right) &=&h_{22}\delta \left( \mathbf{\rho }\right) ;
\label{A9} \\
\left( h_{11}h_{22}-\widehat{h}_{21}\widehat{h}_{12}\right) g_{22}\left(
\mathbf{\rho }\right) &=&h_{11}\delta \left( \mathbf{\rho }\right) \text{.}
\label{A10}
\end{eqnarray}

We present next a detailed calculation of the normal GF, while the associate
GF are obtained similarly. For $g_{11}^{1}$, after substitution of the
matrix elements from Eq. (\ref{a6}), one obtains the following second order
linear differential equation with a source:
\begin{equation}
\left( \left( i\omega +\mu \right) ^{2}-\Pi ^{2}-i\left[ \Pi _{x},\Pi _{y}%
\right] \right) g_{11}^{1}\left( \mathbf{\rho }\right) =\left( i\omega +\mu
\right) \delta \left( \mathbf{\rho }\right) \text{.}  \label{A11}
\end{equation}%
This is written via Laplacian,
\begin{equation}
\widehat{L}=\frac{l^{2}}{2}\left \{ -\frac{\partial ^{2}}{\partial \rho ^{2}}%
-\frac{1}{\rho }\frac{\partial }{\partial \rho }-\frac{1}{\rho ^{2}}\frac{%
\partial ^{2}}{\partial \theta ^{2}}+\frac{i}{2l^{2}}\frac{\partial }{%
\partial \theta }+\frac{\rho ^{2}}{4l^{4}}\right \} \text{,}  \label{A13}
\end{equation}%
as,represented into the form:

\begin{equation}
\left( \left( i\omega +\mu \right) ^{2}-\frac{\omega _{c}^{2}}{2}-\omega
_{c}^{2}\widehat{L}\right) g_{11}^{1}\left( \mathbf{\rho }\right) =\left(
i\omega +\mu \right) \delta \left( \mathbf{\rho }\right) \text{.}
\label{A12}
\end{equation}%
Since the operator $\widehat{L}$ in this equation is rotation invariant, $%
g_{11}^{1}\left( \mathbf{\rho }\right) $ is \ a scalar (independent of the
polar angle). The operator $\widehat{L}$ has the following eigenfunctions
and eigenvalues \cite{Landau}:

\begin{equation}
\epsilon _{n}^{m}=n+\frac{\left \vert m\right \vert +m+1}{2}\text{,}
\label{A15}
\end{equation}%
and {}eigenfunctions%
\begin{equation}
\varphi _{n}^{m}=\frac{1}{l^{1+\left \vert m\right \vert }}\sqrt{\frac{n!}{%
2^{\left \vert m\right \vert }\left( \left \vert m\right \vert +n\right) !}}%
\exp \left[ -\frac{\rho ^{2}}{4l^{2}}\right] \rho ^{\left \vert m\right
\vert }L_{n}^{\left \vert m\right \vert }\left( \frac{\rho ^{2}}{2l^{2}}%
\right) \frac{e^{im\theta }}{\sqrt{2\pi }}\text{.}  \label{A16}
\end{equation}%
Here $n$ and $m$ are integers and $L_{n}^{m}$ are the generalized Laguerre
polynomials.

In specific case of a scalar the azimuthal number $m=0$, and one obtains:
\begin{equation}
\varphi _{n}^{0}=\frac{1}{\sqrt{2\pi }l}\exp \left[ -\frac{\rho ^{2}}{4l^{2}}%
\right] L_{n}\left[ \frac{\rho ^{2}}{2l^{2}}\right] .  \label{A17}
\end{equation}%
Expanding the GF $g_{11}^{1}\left( \mathbf{\rho }\right) $ by series of the
scalar eigenfunctions of the $\widehat{L}$ operator, $g_{11}^{1}\left(
\mathbf{\rho }\right) =\sum_{n}c_{n}^{0}\varphi _{n}^{0}$, and making the
scalar product with $\varphi _{n}^{0}$, one obtains:
\begin{equation}
\int_{\mathbf{\rho }}\varphi _{n^{\prime }}^{0\ast }\sum_{nm}\left[ \left(
i\omega +\mu \right) ^{2}-\omega _{c}^{2}\left( 1+n\right) \right]
c_{n}^{0}\varphi _{n}^{0}=\left( i\omega +\mu \right) \int_{\mathbf{\rho }%
}\varphi _{n^{\prime }}^{0\ast }\left( \mathbf{\rho }\right) \delta \left(
\mathbf{\rho }\right)  \label{A18}
\end{equation}%
Performing the integration, finally

\begin{equation}
g_{11}^{1}\left( \mathbf{\rho }\right) =\frac{i\omega +\mu }{2\pi l^{2}}\exp %
\left[ -\rho ^{2}/4l^{2}\right] \sum_{n=0}\frac{\ L_{n}\left[ \rho
^{2}/2l^{2}\right] }{\left( i\omega +\mu \right) ^{2}-\omega _{c}^{2}\left(
1+n\right) }\text{.}  \label{A19}
\end{equation}

Using the relation Eq.(\ref{A8}), the off diagonal matrix element $%
g_{21}^{1}\left( \mathbf{\rho }\right) $ reads:
\begin{equation}
g_{21}^{1}\left( \mathbf{\rho }\right) =\frac{\omega _{c}}{i\omega +\mu }%
a^{\dagger }g_{11}^{1}\left( \mathbf{\rho }\right) \text{.}  \label{A20}
\end{equation}%
Since
\begin{equation}
a^{\dagger }=\frac{i}{\omega _{c}}e^{i\theta }\left( \frac{\partial }{%
\partial \rho }-\frac{i}{\rho }\frac{\partial }{\partial \theta }+\frac{\rho
}{2l^{2}}\right) ,  \label{A22}
\end{equation}%
using the relation between Laguere polynomials \cite{Gradshtein}, the result
is:

\begin{equation}
g_{21}^{1}\left( \mathbf{\rho }\right) =\frac{i\rho }{2\pi l^{4}}e^{i\theta
}\exp \left[ -\rho ^{2}/4l^{2}\right] \sum_{n=1}\frac{L_{n-1}^{1}\left[ \rho
^{2}/2l^{2}\right] }{\left( i\omega +\mu \right) ^{2}-\omega _{c}^{2}\left(
1+n\right) }\text{.}  \label{A23}
\end{equation}

In order to calculate the next pair of the GF matrix elements, $g_{22}^{1}$
and $g_{12}^{1}$, one has to solve the second Eq.(\ref{A9}). The
corresponding equation is similar,
\begin{equation}
\left( -\omega _{c}^{2}a^{\dagger }a+\left( i\omega +\mu \right) ^{2}\right)
g_{22}^{1}\left( \mathbf{\rho }\right) =\left( i\omega +\mu \right) \delta
\left( \mathbf{\rho }\right) ,  \label{A24}
\end{equation}

\begin{equation}
\left \{ \left( i\omega +\mu \right) ^{2}-\omega _{c}^{2}\widehat{L}\right
\} g_{22}^{1}\left( \mathbf{\rho }\right) \ =\left( i\omega +\mu \right)
\delta \left( \mathbf{\rho }\right) .  \label{A25}
\end{equation}

\bigskip Repeating the procedure this results in
\begin{equation}
g_{22}^{1}\left( \mathbf{\rho }\right) =\frac{i\omega +\mu }{2\pi l^{2}}\exp %
\left[ -\rho ^{2}/4l^{2}\right] \sum_{n=0}\frac{\ L_{n}\left[ \rho
^{2}/2l^{2}\right] }{\left( i\omega +\mu \right) ^{2}-\omega _{c}^{2}n}.
\label{A27}
\end{equation}%
Using the relation $\ g_{12}^{1}\left( \mathbf{\rho }\right) =\frac{\omega
_{c}}{i\omega +\mu }ag_{22}^{1}\left( \mathbf{\rho }\right) $, one obtains
in view of%
\begin{equation*}
a=-\frac{ie^{-i\theta }}{\omega _{c}}\left( -\frac{\partial }{\partial \rho }%
-\frac{i}{\rho }\frac{\partial }{\partial \theta }+\frac{\rho }{2l^{2}}%
\right) ,
\end{equation*}%
\begin{equation}
g_{12}^{1}\left( \mathbf{\rho }\right) =i\frac{ve^{-i\theta }}{2\pi l^{4}}%
\rho \exp \left[ -\frac{\rho ^{2}}{4l^{2}}\right] \sum_{n=1}\frac{L_{n}^{1}%
\left[ \rho ^{2}/2l^{2}\right] }{\left( i\omega +\mu \right) ^{2}-\omega
_{c}^{2}n}.  \label{A28}
\end{equation}%
\  \

The associated GF is calculated in the same manner, replacing matrix
elements as it's presented in Eq.(\ref{a6}). All of the GF are presented in
Eq. (\ref{GF1}),(\ref{GF2}).

\section{Matsubara summations}

The sums over reduced Matsubara frequency $\overline{\omega }_{s}=\pi \left(
2s+1\right) $ in Eq.(\ref{GGG}) read:

\begin{eqnarray}
A_{1}\left[ a,b\right] &=&\sum \nolimits_{s=-\infty }^{\infty }\frac{%
\overline{\omega }_{s}^{2}+\overline{\mu }^{2}}{\left( \left( -i\overline{%
\omega _{s}}+\overline{\mu }\right) ^{2}-\overline{\omega }_{c}^{2}\left(
n+1\right) \right) \left( \left( i\overline{\omega }_{s}+\mu \right) ^{2}-%
\overline{\omega }_{c}^{2}\left( m+1\right) \right) }  \label{B1} \\
&=&\frac{\left( \sqrt{a}-\mu \right) ^{2}\tanh \left( \frac{\sqrt{a}-\mu }{2}%
\right) }{4\sqrt{a}\left( -b+(\sqrt{a}-2\mu \right) ^{2})}+\frac{\left(
\sqrt{b}-\mu \right) ^{2}\tanh \left( \frac{\sqrt{b}-\mu }{2}\right) }{4%
\sqrt{b}\left( -a+\left( \sqrt{b}-2\mu \right) ^{2}\right) };  \notag
\end{eqnarray}

\begin{eqnarray}
A_{2}\left[ a,b\right] &=&\sum \nolimits_{s}\frac{\overline{\omega }_{s}^{2}+%
\overline{\mu }^{2}}{\left( \left( -i\overline{\omega }_{s}+\overline{\mu }%
\right) ^{2}-\overline{\omega }_{c}^{2}n\right) \left( \left( i\overline{%
\omega }_{s}+\overline{\mu }\right) ^{2}-\overline{\omega }_{c}^{2}m\
\right) }  \label{B2} \\
&=&\frac{\left( \sqrt{a}+\mu \right) ^{2}\tanh \left( \frac{\sqrt{a}+\mu }{2}%
\right) }{4\sqrt{a}\left( -b+(\sqrt{a}+2\mu \right) ^{2})}+\frac{\left(
\sqrt{b}+\mu \right) ^{2}\tanh \left( \frac{\sqrt{b}+\mu }{2}\right) }{4%
\sqrt{b}\left( -a+\left( \sqrt{b}+2\mu \right) ^{2}\right) };  \notag
\end{eqnarray}%
\begin{eqnarray}
B_{1}\left[ a,b\right] &=&\sum \nolimits_{s}\frac{n}{\left( \left( -i%
\overline{\omega }_{s}+\overline{\mu }\right) ^{2}-\overline{\omega }%
_{s}^{2}\left( n+1\right) \right) \left[ \left( \left( i\overline{\omega }%
_{s}+\overline{\mu }\right) ^{2}-\overline{\omega }_{c}^{2}m\right) \right] }
\label{B3} \\
&=&-\frac{\tanh \left( \frac{\sqrt{a}-\overline{\mu }}{2}\right) }{4\sqrt{a}%
\left( -b+(\sqrt{a}-2\overline{\mu }\right) ^{2})}-\frac{\tanh \left( \frac{%
\sqrt{b}-\overline{\mu }}{2}\right) }{4\sqrt{b}\left( -a+\left( \sqrt{b}-2%
\overline{\mu }\right) ^{2}\right) };  \notag
\end{eqnarray}%
and%
\begin{eqnarray}
B_{2}\left[ a,b\right] &=&\sum \nolimits_{s}\frac{m}{\left( \left( -i%
\overline{\omega }_{s}+\overline{\mu }\right) ^{2}-\overline{\omega }%
_{c}^{2}n\right) \left( \left( i\overline{\omega }_{s}+\overline{\mu }%
\right) ^{2}-\overline{\omega }_{c}^{2}\left( m+1\right) \right) }
\label{B4} \\
&=&-\frac{\tanh \left[ \frac{\sqrt{a}+\overline{\mu }}{2}\right] }{4\sqrt{a}%
\left( -b+(\sqrt{a}+2\overline{\mu }\right) ^{2})}-\frac{\tanh \left[ \frac{%
\sqrt{b}+\overline{\mu }}{2}\right] }{4\sqrt{b}\left( -a+\left( \sqrt{b}+2%
\overline{\mu }\right) ^{2}\right) }.  \notag
\end{eqnarray}%
Functions $A\left[ a,b\right] $ and $B\left[ a,b\right] $ in the Eq.(\ref%
{Umn}) are subsequently composed as:

\begin{equation}
A\left[ a,b\right] =A_{1}\left[ a,b\right] +A_{2}\left[ a,b\right] ;\text{ \
\  \  \  \  \ }B\left[ a,b\right] =B_{1}\left[ a,b\right] +B_{2}\left[ a,b\right]
\text{.}  \label{B5}
\end{equation}

\section{\protect \bigskip Zeeman Effect}

\subsection{The Zeeman term in Gorkov equations}

In the case of the WSM Hamiltonian containing the Zeeman term, Eq.(\ref{H_Z}%
), the Gor'kov equations for normal Green Function at criticality reads,
\begin{equation}
\frac{\partial G_{\gamma \kappa }^{st}\left( X,X^{\prime }\right) }{\partial
\tau }=\ i\sigma _{\gamma \beta }^{i}\partial _{r}G_{\beta \kappa
}^{st}\left( X,X^{\prime }\right) +\mu G_{\gamma \kappa }^{st}\left(
X,X^{\prime }\right) +g_{L}\mu _{B}H\tau _{st^{\prime }}^{z}G_{\gamma \kappa
}^{t^{\prime }t}\left( X,X^{\prime }\right) -\delta ^{\gamma \kappa }\delta
^{ts}\delta \left( X-X^{\prime }\right) ,  \label{C1}
\end{equation}%
while the equation for the anomalous average becomes:

\begin{equation}
\frac{\partial F_{\gamma \kappa }^{st+}\left( X,X^{\prime }\right) }{%
\partial \tau }=iv\sigma _{\alpha \gamma }^{i}\nabla _{r}^{i}F_{\alpha
\kappa }^{st+}\left( X,X^{\prime }\right) -\mu F_{\gamma \kappa
}^{+st}\left( X,X^{\prime }\right) -\frac{g^{2}}{4}\varepsilon
^{s_{1}s_{2}}F_{\alpha \gamma }^{+s_{1}s_{2}}\left( X,X\right) \varepsilon
^{s_{3}s}G_{\alpha \kappa }^{s_{3}t}-g_{L}\mu _{B}H\tau _{st^{\prime
}}^{z}F_{\gamma \kappa }^{t^{\prime }t+}\left( X,X^{\prime }\right) \text{.}
\label{C2}
\end{equation}%
Number of GF in this case is doubled, although due to symmetry for singlet
pairing solution one observes that $G_{\gamma \kappa }^{\uparrow \downarrow
}=G_{\gamma \kappa }^{\downarrow \uparrow }=$ $F_{\gamma \kappa }^{+\uparrow
\uparrow }=F_{\gamma \kappa }^{+\downarrow \downarrow }=0$.

Self - consistent equation for the gap function is,

\begin{equation}
\Delta _{\beta \kappa }^{\ast }\left( \mathbf{r}\right) =-\frac{g^{2}}{4}%
\int_{\mathbf{r}^{\prime }}\left( G_{\beta \gamma }^{2\downarrow \downarrow
}\left( \mathbf{r}^{\prime }\mathbf{,r}\right) \Delta _{\alpha \gamma
}^{\ast }\left( \mathbf{r}^{\prime }\right) G_{\alpha \kappa }^{1\uparrow
\uparrow }\left( \mathbf{r,r}^{\prime }\right) +G_{\beta \gamma }^{2\uparrow
\uparrow +}\left( \mathbf{r}^{\prime }\mathbf{,r}\right) \Delta _{\alpha
\gamma }^{\ast }\left( \mathbf{r}^{\prime }\right) G_{\alpha \kappa
}^{1\downarrow \downarrow }\left( \mathbf{r,r}^{\prime }\right) \right) ,
\label{C3}
\end{equation}%
while the GF in magnetic field are
\begin{eqnarray}
G_{\beta \kappa }^{ss1}\left( \mathbf{r},\mathbf{r}^{\prime }\right) &=&\exp %
\left[ -i\frac{xy^{\prime }-yx^{\prime }}{2l^{2}}\right] g_{\beta \kappa
}^{ss1}\left( \mathbf{r-r}^{\prime }\right) ;  \label{C4} \\
G_{\beta \kappa }^{ss2}\left( \mathbf{r}^{\prime },\mathbf{r}\right) &=&\exp %
\left[ -i\frac{xy^{\prime }-yx^{\prime }}{2l^{2}}\right] g_{\beta \kappa
}^{ss2}\left( \mathbf{r}^{\prime }\mathbf{-r}\right) \text{.}  \notag
\end{eqnarray}%
here $s=\uparrow ,\downarrow .$

Substituting Eq.(\ref{C4}) into Eq.(\ref{C3}) and using the singlet
assumption, $\Delta _{\alpha \gamma }^{\ast }\left( \mathbf{r}\right)
=\Delta \left( \mathbf{r}\right) \sigma _{\alpha \gamma }^{x}$, one obtains
Eq.(\ref{kkk}), and after the angle integration Eq.(\ref{gapeqangle}) with
the only difference being the modified function $S$:

\begin{equation}
S_{Z}\left( \rho ,\omega \right) =\left(
\begin{array}{c}
g_{21}^{2\downarrow \downarrow }\left( \mathbf{-\rho }\right) \
g_{21}^{1\uparrow \uparrow }\left( \mathbf{\rho }\right) +\
g_{22}^{2\uparrow \uparrow }\left( \mathbf{-\rho }\right) \
g_{11}^{1\downarrow \downarrow }\left( \mathbf{\rho }\right) +\
g_{22}^{2\downarrow \downarrow }\left( \mathbf{-\rho }\right) \
g_{11}^{1\uparrow \uparrow }\left( \mathbf{\rho }\right) +\
g_{21}^{2\uparrow \uparrow }\left( \mathbf{-\rho }\right) \
g_{21}^{1\downarrow \downarrow }\left( \mathbf{\rho }\right) \\
\ g_{11}^{2\downarrow \downarrow }\left( \mathbf{-\rho }\right) \
g_{22}^{1\uparrow \uparrow }\left( \mathbf{\rho }\right) +\
g_{11}^{2\uparrow \uparrow }\left( \mathbf{-\rho }\right) \
g_{22}^{1\downarrow \downarrow }\left( \mathbf{\rho }\right) +\
g_{12}^{2\downarrow \downarrow }\left( \mathbf{-\rho }\right) \
g_{12}^{1\uparrow \uparrow }\left( \mathbf{\rho }\right) +g_{12}^{2\uparrow
\uparrow }\left( \mathbf{-\rho }\right) \ g_{12}^{1\downarrow \downarrow
}\left( \mathbf{\rho }\right)%
\end{array}%
\right) .  \label{C5}
\end{equation}

\subsection{\protect \bigskip Calculation of the GF}

Calculation of the GF is performed along the lines described in Appendix A.
In this case however we get two separate equations for each GF with
different spin projections. The equations for first GF are:

\begin{eqnarray}
i\omega G_{\gamma \kappa }^{1\uparrow \uparrow }\left( \mathbf{r,r}^{\prime
}\right) \ -i\sigma _{\gamma \beta }^{i}\partial _{r}G_{\beta \kappa
}^{1\uparrow \uparrow }\left( \mathbf{r,r}^{\prime }\right) +\left( \mu
+g_{L}\mu _{B}H\right) G_{\gamma \kappa }^{1\uparrow \uparrow }\left(
\mathbf{r,r}^{\prime }\right) &=&\delta ^{\gamma \kappa }\delta \left(
\mathbf{r-r}^{\prime }\right) ;  \label{C6} \\
i\omega G_{\gamma \kappa }^{1\downarrow \downarrow }\left( \mathbf{r,r}%
^{\prime }\right) \ -i\sigma _{\gamma \beta }^{i}\partial _{r}G_{\beta
\kappa }^{1\downarrow \downarrow }\left( \mathbf{r,r}^{\prime }\right)
+\left( \mu -g_{L}\mu _{B}H\right) G_{\gamma \kappa }^{1\downarrow
\downarrow }\left( \mathbf{r,r}^{\prime }\right) &=&\delta ^{\gamma \kappa
}\delta \left( \mathbf{r-r}^{\prime }\right) .  \notag
\end{eqnarray}%
Therefore the solution coincides with that of the GF Eq.(\ref{GF1}) for two
different values of the chemical potential. The result is

\begin{eqnarray}
g_{11}^{1\uparrow \uparrow ,\downarrow \downarrow }\left( \mathbf{\rho }%
\right) &=&\frac{\left( i\omega +\mu \pm g_{L}\mu _{B}H\right) }{2\pi l^{2}}%
\exp \left[ -\frac{\rho ^{2}}{4l^{2}}\right] \sum_{n=0}\frac{\ L_{n}\left[
\rho ^{2}/2l^{2}\right] }{\left( i\omega +\mu \pm g_{L}\mu _{B}H\right)
^{2}-\omega _{c}^{2}\left( 1+n\right) };  \label{C7} \\
g_{21}^{1\uparrow \uparrow ,\downarrow \downarrow }\left( \mathbf{\rho }%
\right) &=&\frac{iv\rho e^{i\theta }}{2\pi l^{4}}\exp \left[ -\frac{\rho ^{2}%
}{4l^{2}}\right] \sum_{n=1}\frac{L_{n-1}^{1}\left[ \rho ^{2}/2l^{2}\right] }{%
\left( i\omega +\mu \pm g_{L}\mu _{B}H\right) ^{2}-\omega _{c}^{2}\left(
1+n\right) };  \notag \\
g_{22}^{1\uparrow \uparrow ,\downarrow \downarrow }\left( \mathbf{\rho }%
\right) &=&\frac{\left( i\omega +\mu \pm g_{L}\mu _{B}H\right) }{2\pi l^{2}}%
\exp \left[ -\frac{\rho ^{2}}{4l^{2}}\right] \sum_{n=0}\frac{\ L_{n}\left[
\rho ^{2}/2l^{2}\right] }{\left( i\omega +\mu \pm g_{L}\mu _{B}H\right)
^{2}-\omega _{c}^{2}n};  \notag \\
g_{12}^{1\uparrow \uparrow ,\downarrow \downarrow }\left( \mathbf{\rho }%
\right) &=&\frac{iv\rho e^{-i\theta }}{2\pi l^{4}}\exp \left[ -\frac{\rho
^{2}}{4l^{2}}\right] \sum_{n=1}\frac{L_{n}^{1}\left[ \rho ^{2}/2l^{2}\right]
}{\left[ \left( i\omega +\mu \pm \mu _{Z}H\right) ^{2}-\omega _{c}^{2}n\ %
\right] }.  \notag
\end{eqnarray}

Similarly for the second set of GF:%
\begin{eqnarray}
g_{11}^{2\uparrow \uparrow ,\downarrow \downarrow }\left( -\mathbf{\rho }%
\right) &=&\frac{\left( -i\omega +\mu \pm g_{L}\mu _{B}H\right) }{2\pi l^{2}}%
\exp \left[ -\frac{\rho ^{2}}{4l^{2}}\right] \sum_{n=0}\frac{L_{n}\left[
\rho ^{2}/2l^{2}\right] }{\left( -i\omega +\mu \pm g_{L}\mu _{B}H\right)
^{2}-\omega _{c}^{2}n};  \label{C8} \\
g_{12}^{2\uparrow \uparrow ,\downarrow \downarrow }\left( -\mathbf{\rho }%
\right) &=&\frac{iv\rho e^{i\theta }}{2\pi l^{4}}\exp \left[ -\frac{\rho ^{2}%
}{4l^{2}}\right] \sum_{n=1}^{\infty }\frac{L_{n-1}^{1}\left[ \rho ^{2}/2l^{2}%
\right] }{\left( -i\omega +\mu \pm g_{L}\mu _{B}H\right) ^{2}-\omega
_{c}^{2}\left( n+1\right) };  \notag \\
g_{21}^{2\uparrow \uparrow ,\downarrow \downarrow }\left( -\mathbf{\rho }%
\right) &=&-\frac{iv\rho e^{-i\theta }}{2\pi l^{4}}\exp \left[ -\frac{\rho
^{2}}{4l^{2}}\right] \sum_{n=1}^{\infty }\frac{L_{n}^{1}\left[ \rho
^{2}/2l^{2}\right] }{\left( -i\omega +\mu \pm g_{L}\mu _{B}H\right)
^{2}-\omega _{c}^{2}n};  \notag \\
g_{22}^{2\uparrow \uparrow ,\downarrow \downarrow }\left( -\mathbf{\rho }%
\right) &=&\frac{\left( -i\omega +\mu \pm g_{L}\mu _{B}H\right) }{2\pi l^{2}}%
\exp \left[ -\frac{\rho ^{2}}{4l^{2}}\right] \sum_{n=0}^{\infty }\frac{L_{n}%
\left[ \rho ^{2}/2l^{2}\right] }{\left( -i\omega +\mu \pm g_{L}\mu
_{B}H\right) ^{2}-\omega _{c}^{2}\left( n+1\right) }  \notag
\end{eqnarray}

\bigskip

\section{$\protect \bigskip $Generalization to 3D}

\subsection{Density of states for a film in zero magnetic field}

Using the dispersion law in the form,
\begin{equation}
\varepsilon =\sqrt{v^{2}\left( p_{x}^{2}+p_{y}^{2}\right) +v_{z}^{2}p_{z}^{2}%
},  \label{D1}
\end{equation}%
one obtains for the density of electrons for the bulk anisotropic sample,

\begin{equation}
n=\frac{1}{\left( 2\pi \right) ^{3}\hbar ^{3}}\int_{p}\Theta \left(
\varepsilon \left[ p\right] -\mu \right) =\frac{\mu ^{3}}{6\pi
^{2}v_{z}c_{x}^{2}\hbar ^{3}},  \label{D2}
\end{equation}%
while the density of electron states

\begin{equation}
D\left( \mu \right) =\frac{\mu ^{2}}{2\pi ^{2}v_{z}v^{2}\hbar ^{3}}.
\label{D3}
\end{equation}

In films of thickness $d$ the quantization of the momentum along axes $z$ is
important and the density of the electrons reads,

\begin{equation}
n\left[ \mu \right] =\frac{N}{Ad}=\frac{1}{\left( 2\pi \right) ^{2}\hbar ^{2}%
}\frac{1}{2d}\int_{\mathbf{p}}\sum \limits_{M}\Theta \left( \varepsilon %
\left[ \mathbf{p},M\right] -\mu \right) ,  \label{D4}
\end{equation}%
where $\varepsilon ^{2}\left[ p,M\right] =v^{2}\left(
p_{x}^{2}+p_{y}^{2}\right) +v_{z}^{2}\left( \pi \hbar M/d\right)
^{2}=v^{2}p^{2}+v_{z}^{2}\left( \pi \hbar M/d\right) ^{2},$ and the chemical
potential is $\mu =\sqrt{v^{2}u+v_{z}^{2}\left( \pi \hbar M/d\right) ^{2}}$.
The density of states in this case is
\begin{equation}
D\left( \mu \right) =\frac{dn}{d\mu }=\frac{1}{8\pi \hbar ^{2}d}\sum
\limits_{M:\mu >\mu _{M}}\frac{2\mu }{v^{2}}=\frac{\mu }{4\pi \hbar
^{2}dv^{2}}F\left[ \mu \right] \text{.}  \label{D5}
\end{equation}%
Here $\  \mu _{M}\equiv \frac{\pi \hbar v_{z}}{d}\left \vert M\right \vert ,M%
\left[ \mu \right] =\frac{d\mu _{M}}{\pi \hbar v_{z}}$ and $F\left[ \mu %
\right] $ is the step-like function $F=2n$ in the interval $n\pi \hbar
v_{z}/d<\mu <\left( n+1\right) \pi \hbar v_{z}/d\ ,n=1,2,3,...$

\subsection{Green's functions in 3D}

In this Appendix the normal state Green's functions for 3D are calculated.
In the matrix form the equations (\ref{normalg1}), (\ref{n4ac}) read:
\begin{equation}
\widehat{h}^{a}g^{a}\left( \mathbf{\rho }\right) \ =\delta \left( \mathbf{%
\rho }\right) ;  \label{D6}
\end{equation}%
where $a=1,2,$with 3D matrix operators%
\begin{equation}
\widehat{h}^{1}=i\omega +\mu -\mathbf{\Pi }\cdot \mathbf{\sigma -}%
v_{z}p_{z}\sigma ^{z}\mathbf{;}\widehat{h}^{2}=-i\omega +\mu +\mathbf{\Pi
\cdot \sigma }^{t}+v_{z}p_{z}\sigma ^{z}  \label{D7}
\end{equation}%
Substituting $\widehat{h}^{1}$ and $\widehat{h}^{2}$ into Eq.(\ref{C6}) and
solving set of eight equations in the manner similar to that described in
Appendix A, one obtains the first set of GF
\begin{eqnarray}
g_{11}^{1}\left( \mathbf{\rho ,}p_{z}\right) &=&\frac{v_{z}p_{z}+i\omega
+\mu }{2\pi l^{2}}\exp \left[ -\frac{\rho ^{2}}{4l^{2}}\right] \sum_{n=0}%
\frac{L_{n}\left[ \rho ^{2}/2l^{2}\right] }{\left( i\omega +\mu \right)
^{2}-v_{z}^{2}p_{z}^{2}-\omega _{c}^{2}\left( n+1\right) };  \label{D8} \\
g_{12}^{1}\left( \mathbf{\rho },p_{z}\right) &=&-\frac{i\rho e^{-i\theta }}{%
2\pi l^{4}}\exp \left[ -\frac{\rho ^{2}}{4l^{2}}\right] \  \  \sum_{n=1}\frac{%
L_{n}^{1}\left[ \rho ^{2}/2l^{2}\right] \ }{\left( i\omega +\mu \right)
^{2}-v_{z}^{2}p_{z}^{2}-\omega _{c}^{2}n\ };  \notag \\
g_{22}^{1}\left( \mathbf{\rho },p_{z}\right) &=&\frac{-v_{z}p_{z}+i\omega
+\mu }{2\pi l^{2}}\exp \left[ -\frac{\rho ^{2}}{4l^{2}}\right] \sum_{n=0}%
\frac{L_{n}\left[ \rho ^{2}/2l^{2}\right] }{\left( i\omega +\mu \right)
^{2}-v_{z}^{2}p_{z}^{2}-\omega _{c}^{2}n\ };  \notag \\
g_{21}^{1}\left( \mathbf{\rho },p_{z}\right) &=&-\frac{ie^{i\theta }\rho }{%
2\pi l^{4}}\exp \left[ -\frac{\rho ^{2}}{4l^{2}}\right] \sum_{n=1}\frac{%
L_{n-1}^{1}\left[ \rho ^{2}/2l^{2}\right] \ }{\left( i\omega +\mu \right)
^{2}-v_{z}^{2}p_{z}^{2}-\omega _{c}^{2}\left( n+1\right) }\text{,}  \notag
\end{eqnarray}%
and the second set,

\begin{eqnarray}
g_{11}^{2}\left( -\mathbf{\rho ,-}p_{z}\right) &=&\frac{-v_{z}p_{z}+i\omega
+\mu }{2\pi l^{2}}\exp \left[ -\frac{\rho ^{2}}{4l^{2}}\right] \sum_{n=0}%
\frac{L_{n}\left[ \rho ^{2}/2l^{2}\right] }{\left( i\omega +\mu \right)
^{2}-v_{z}^{2}p_{z}^{2}-\omega _{c}^{2}\left( 1+n\right) };  \label{D9} \\
g_{12}^{2}\left( -\mathbf{\rho },-p_{z}\right) &=&\frac{ie^{i\theta }\rho }{%
2\pi l^{4}}\exp \left[ -\frac{\rho ^{2}}{4l^{2}}\right] \  \  \sum_{n=1}\frac{%
L_{n}^{1}\left[ \rho ^{2}/2l^{2}\right] \ }{\left( i\omega +\mu \right)
^{2}-v_{z}^{2}p_{z}^{2}-\omega _{c}^{2}n\ };  \notag \\
g_{22}^{2}\left( -\mathbf{\rho },-p_{z}\right) &=&\frac{v_{z}p_{z}+i\omega
+\mu }{2\pi l^{2}}\exp \left[ -\frac{\rho ^{2}}{4l^{2}}\right] \sum_{n=0}%
\frac{L_{n}\left[ \rho ^{2}/2l^{2}\right] }{\left( i\omega +\mu \right)
^{2}-v_{z}^{2}p_{z}^{2}-\omega _{c}^{2}n\ };  \notag \\
g_{21}^{2}\left( -\mathbf{\rho },-p_{z}\right) &=&\frac{ie^{-i\theta }\rho }{%
2\pi l^{4}}\exp \left[ -\frac{\rho ^{2}}{4l^{2}}\right] \sum_{n=1}\frac{%
L_{n-1}^{1}\left[ \rho ^{2}/2l^{2}\right] \ }{\left( i\omega +\mu \right)
^{2}-v_{z}^{2}p_{z}^{2}-\omega _{c}^{2}\left( 1+n\right) }\text{.}  \notag
\end{eqnarray}

These functions allow to solve exactly the gap equation.

\subsection{Solution of the gap equation in 3D}

The gap equation in 3D takes a form:
\begin{equation}
\Delta \left( \mathbf{r}\right) =\frac{g^{2}T}{2}\sum \nolimits_{\omega
}\int_{\mathbf{r}^{\prime }}\exp \left[ -i\frac{xy^{\prime }-yx^{\prime }}{%
l^{2}}\right] \Delta ^{\ast }\left( \mathbf{r}^{\prime }\right) \left[
\begin{array}{c}
g_{22}^{2}\left( \mathbf{-\rho ,}-p_{z}\right) g_{11}^{1}\left( \mathbf{\rho
,}p_{z}\right) +g_{11}^{2}\left( \mathbf{-\rho ,}-p_{z}\right)
g_{22}^{1}\left( \mathbf{\rho ,}p_{z}\right) \\
+g_{12}^{2}\left( \mathbf{-\rho ,}-p_{z}\right) g_{12}^{1}\left( \mathbf{%
\rho ,}p_{z}\right) +g_{21}^{2}\left( \mathbf{-\rho ,}-p_{z}\right)
g_{21}^{1}\left( \mathbf{\rho ,}p_{z}\right)%
\end{array}%
\right] \text{,}  \label{D10}
\end{equation}%
where $\mathbf{\rho =r-r}^{\prime }$, $\mathbf{r,r}^{\prime }$ are vectors
in the $x-y$ plane. Substituting the Ansatz for the gap function, Eq.(\ref%
{deltaAnsatz}), and GF, Eqs.(\ref{D8}) and (\ref{D9}), into Eq.(\ref{D10}),
and performing integration over the angle as in 2D case, one obtains the
equation (using notation $u=\rho ^{2}/2l^{2}$):

\begin{equation}
\frac{2}{g^{2}}=\frac{1}{2\pi l^{2}}\sum \nolimits_{\omega
,p_{z}}\int_{u}e^{-2u}S\left( u,p_{z},\omega \right) \text{.}  \label{D11}
\end{equation}%
Here

\begin{equation}
S\left( u,p_{z},\omega \right) =%
\begin{array}{c}
\sum \limits_{n,m=0}\left \{ \frac{\left( \omega ^{2}+\left( \mu
-v_{z}p_{z}\right) ^{2}\right) L_{n}\left[ u\right] L_{m}\left[ u\right] }{%
\left( \left( -i\omega +\mu \right) ^{2}-v_{z}^{2}p_{z}^{2}-\omega
_{c}^{2}\left( n+1\right) \right) \left( \left( i\omega +\mu \right)
^{2}-v_{z}^{2}p_{z}^{2}-\omega _{c}^{2}\left( m+1\right) \right) }%
+\sum_{n,m=0}\frac{\left( \omega ^{2}+\left( \mu +v_{z}p_{z}\right)
^{2}\right) L_{n}\left[ u\right] L_{m}\left[ u\right] }{\left( \left(
-i\omega +\mu \right) ^{2}-v_{z}^{2}p_{z}^{2}-\omega _{c}^{2}n\right) \left(
\left( i\omega +\mu \right) ^{2}-v_{z}^{2}p_{z}^{2}-\omega _{c}^{2}m\right) }%
\right \} \\
+\sum \limits_{n,m=1}\left \{ \frac{\omega _{c}^{2}uL_{n-1}^{1}\left[ u%
\right] L_{m}^{1}\left[ u\right] }{\left( \left( -i\omega +\mu \right)
^{2}-v_{z}^{2}p_{z}^{2}-\omega _{c}^{2}\left( n+1\right) \right) \left(
\left( i\omega +\mu \right) ^{2}-v_{z}^{2}p_{z}^{2}-\omega _{c}^{2}m\right) }%
+\frac{\omega _{c}^{2}uL_{n}^{1}\left[ u\right] L_{m-1}^{1}\left[ u\right] }{%
\left( \left( -i\omega +\mu \right) ^{2}-v_{z}^{2}p_{z}^{2}-\omega
_{c}^{2}n\right) \left( \left( i\omega +\mu \right)
^{2}-v_{z}^{2}p_{z}^{2}-\omega _{c}^{2}\left( m+1\right) \right) }\right \}%
\end{array}%
.  \label{D12}
\end{equation}

After integration over $u$ it is written as a double sum:

\begin{equation}
\frac{1}{\lambda }=\frac{\zeta \overline{\omega }_{c}^{2}}{4\mu ^{2}}\sum
\nolimits_{s,M}\left \{ \sum \limits_{n,m=1,s}\frac{\left( m+n\right) !f%
\left[ n\right] f\left[ m\right] }{2^{m+n+1}m!n!}S_{1}+\sum \limits_{n=1,s}%
\frac{f\left[ n\right] f\left[ 0\right] }{2^{n}}S_{2}+\frac{f\left[ 0\right]
^{2}}{2}\sum \limits_{s}S_{3}\right \} ,  \label{D13}
\end{equation}%
where

\begin{eqnarray}
S_{1} &=&%
\begin{array}{c}
\frac{\omega _{s}^{2}+\mu ^{2}+\left( \zeta M\right) ^{2}}{\left( \left(
-i\omega _{s}+\mu \right) ^{2}-\left( \zeta M\right) ^{2}-\omega
_{c}^{2}\left( n+1\right) \right) \left( \left( i\omega _{s}+\mu \right)
^{2}-\left( \zeta M\right) ^{2}-\omega _{c}^{2}\left( m+1\right) \right) }+%
\frac{\omega _{s}^{2}+\mu ^{2}+\left( \zeta M\right) ^{2}}{\left( \left(
-i\omega +\mu \right) ^{2}-\left( \zeta M\right) ^{2}-\omega
_{c}^{2}n\right) \left( \left( i\omega +\mu \right) ^{2}-\left( \zeta
M\right) ^{2}-\omega _{c}^{2}m\right) } \\
+\frac{n\overline{\omega }_{c}^{2}}{\left( \left( -i\omega _{s}+\mu \right)
^{2}-\left( \zeta M\right) ^{2}-\omega _{c}^{2}\left( n+1\right) \right)
\left( \left( i\omega +\mu \right) ^{2}-v_{z}^{2}p_{z}^{2}-\omega
_{c}^{2}m\right) }+\frac{m\overline{\omega }_{c}^{2}}{\left( \left( -i\omega
_{s}+\mu \right) ^{2}-\left( \zeta M\right) ^{2}-\omega _{c}^{2}n\right)
\left( \left( i\omega _{s}+\mu \right) ^{2}-\left( \zeta M\right)
^{2}-\omega _{c}^{2}\left( m+1\right) \right) }%
\end{array}%
;  \label{d14} \\
S_{2} &=&\frac{\omega _{s}^{2}+\mu ^{2}+\left( \zeta M\right) ^{2}}{\left(
\left( -i\omega _{s}+\mu \right) ^{2}-\left( \zeta M\right) ^{2}-\omega
_{c}^{2}\left( n+1\right) \right) \left( \left( i\omega _{s}+\mu \right)
^{2}-\left( \zeta M\right) ^{2}-\omega _{c}^{2}\right) }  \notag \\
&&+\frac{\omega ^{2}+\mu ^{2}+\left( \zeta M\right) ^{2}}{\left( \left(
-i\omega _{s}+\mu \right) ^{2}-\left( \zeta M\right) ^{2}-\omega
_{c}^{2}n\right) \left( \left( i\omega _{s}+\mu \right) ^{2}-\left( \zeta
M\right) ^{2}\right) \ };  \notag \\
S_{3} &=&\frac{\left( \omega _{s}^{2}+\mu ^{2}+\left( \zeta M\right)
^{2}\right) }{\left( \left( -i\omega _{s}+\mu \right) ^{2}-\left( \zeta
M\right) ^{2}-\omega _{c}^{2}\right) \left( \left( i\omega _{s}+\mu \right)
^{2}-\left( \zeta M\right) ^{2}-\omega _{c}^{2}\right) }  \notag \\
&&+\frac{\left( \omega _{s}^{2}+\mu ^{2}+\left( \zeta M\right) ^{2}\right) }{%
\left( \left( -i\omega _{s}+\mu \right) ^{2}-\left( \zeta M\right)
^{2}\right) \left( \left( i\omega _{s}+\mu \right) ^{2}-\left( \zeta
M\right) ^{2}\right) }  \notag
\end{eqnarray}

The abbreviations are as in 2D and in addition $v_{z}\rightarrow v_{z}/T$.
For 3D, after performing summation on Matsubara frequencies, one finally
obtains,

\begin{equation}
\frac{1}{\lambda }=\frac{\zeta \overline{\omega }_{c}^{2}}{4\mu ^{2}}%
\sum \limits_{M>0}\left \{ \sum \limits_{n,m}\frac{\left( m+n\right) !}{%
2^{m+n+1}}\frac{f\left[ n\right] f\left[ m\right] }{m!n!}s_{nmM}+\sum%
\limits_{n}\frac{f\left[ n\right] f\left[ 0\right] }{2^{n}}s_{nM}+\frac{f%
\left[ 0\right] ^{2}}{2}s_{M}\right \} .
\end{equation}%
The summands are,

\begin{eqnarray}
s_{nmM} &=&A\left[ \overline{\omega }_{c}^{2}\left( n+1\right) +\left( \zeta
M\right) ^{2},\overline{\omega }_{c}^{2}\left( m+1\right) +\left( \zeta
M\right) ^{2}\right] +A\left[ \overline{\omega }_{c}^{2}n+\left( \zeta
M\right) ^{2},\overline{\omega }_{c}^{2}m+\left( \zeta M\right) ^{2}\right]
\label{D14} \\
&&+\left( \mu ^{2}+\left( \zeta M\right) ^{2}\right) \left(
\begin{array}{c}
B\left[ \overline{\omega }_{c}^{2}\left( n+1\right) +\left( \zeta M\right)
^{2},\overline{\omega }_{c}^{2}\left( m+1\right) +\left( \zeta M\right) ^{2}%
\right]  \\
+B\left[ \overline{\omega }_{c}^{2}n+\left( \zeta M\right) ^{2},\overline{%
\omega }_{c}^{2}m+\left( \zeta M\right) ^{2}\right]
\end{array}%
\right)   \notag \\
&&+n\overline{\omega }_{c}^{2}B\left[ \overline{\omega }_{c}^{2}\left(
n+1\right) +\left( \zeta M\right) ^{2},\overline{\omega }_{c}^{2}m+\left(
\zeta M\right) ^{2}\  \right] +m\overline{\omega }_{c}^{2}B\left[ \overline{%
\omega }_{c}^{2}n+\left( \zeta M\right) ^{2},\overline{\omega }%
_{c}^{2}\left( m+1\right) +\left( \zeta M\right) ^{2}\right] ,  \notag
\end{eqnarray}%
\begin{eqnarray}
s_{nM} &=&A\left[ \overline{\omega }_{c}^{2}\left( n+1\right) +\left( \zeta
M\right) ^{2},\overline{\omega }_{c}^{2}+\left( \zeta M\right) ^{2}\right] +A%
\left[ \overline{\omega }_{c}^{2}n+\left( \zeta M\right) ^{2},\left( \zeta
M\right) ^{2}\right]  \\
&&+\left( \mu ^{2}+\left( \zeta M\right) ^{2}\right) B\left[ \overline{%
\omega }_{c}^{2}\left( n+1\right) +\left( \zeta M\right) ^{2},\overline{%
\omega }_{c}^{2}+\left( \zeta M\right) ^{2}\right] +\left( \mu ^{2}+\left(
\zeta M\right) ^{2}\right) G\left[ \overline{\omega }_{c}^{2}n+\left( \zeta
M\right) ^{2},\left( \zeta M\right) ^{2}\right] ,  \notag
\end{eqnarray}%
and

\begin{eqnarray}
s_{M} &=&A\left[ \overline{\omega }_{c}^{2}+\left( \zeta M\right) ^{2},%
\overline{\omega }_{c}^{2}+\left( \zeta M\right) ^{2}\right] +A\left[ \left(
\zeta M\right) ^{2},\left( \zeta M\right) ^{2}\right]  \label{d15} \\
&&+\left( \mu ^{2}+\left( \zeta M\right) ^{2}\right) B\left[ \overline{%
\omega }_{c}^{2}+\left( \zeta M\right) ^{2},\overline{\omega }%
_{c}^{2}+\left( \zeta M\right) ^{2}\right] +\left( \mu ^{2}+\left( \zeta
M\right) ^{2}\right) B\left[ \left( \zeta M\right) ^{2},\left( \zeta
M\right) ^{2}\right] \text{,}  \notag
\end{eqnarray}%
with functions $A$ and $B$ given in Appendix B.

\end{document}